\documentclass[journal,draftclsnofoot,onecolumn,12pt]{IEEEtran}
\pdfoutput=1
\usepackage{amsthm,amssymb,graphicx,multirow,amsmath,color,amsfonts}%,ulem}
\usepackage[caption=false,font=footnotesize]{subfig}
\usepackage[update,prepend]{epstopdf}
\usepackage[noadjust]{cite}
\usepackage[latin1]{inputenc}
\usepackage{tikz}
\usepackage{bbm} % for \mathbbm{1}
\usepackage{pdfpages}
\usepackage{multirow}
\usepackage{comment}
\usepackage{algorithm}
\usepackage{url}
\usepackage{algpseudocode}
\def\BSTATE{\STATE\hskip-\ALG@thistlm}
\makeatother

\def\nb0{{\mathbf{0}}}
\def\nb1{{\mathbf{1}}}

\def\ncalA{{\mathcal{A}}}

\def\ncalS{{\mathcal{S}}}

\newtheorem{lemma}{Lemma}

\newtheorem{theorem}{Theorem}

\newtheorem{remark}{Remark}

\def\P{\mathbb{P}}

\def\g{\left.\right|}

\def\a{\overset{(a)}{=}}
\def\b{\overset{(b)}{=}}

\allowdisplaybreaks % Allows breaking of eqnarray over multiple pages (avoids unnecessary blanks in the document before eqnarray)

\begin{document}
%\pagenumbering{gobble}
\graphicspath{{./Figures/}}
\title{
A Reinforcement Learning Framework for Optimizing Age-of-Information in RF-powered Communication Systems %A Generalized Framework for Spatially Clustered RF-powered IoT Network
%Modeling and Analysis of RF-powered IoT Network using Poisson Cluster Process
%Performance Analysis of IoT RF-Powered by Secrecy Wireless Networks 
}
\author{
Mohamed A. Abd-Elmagid, Harpreet S. Dhillon, and Nikolaos Pappas
\thanks{M. A. Abd-Elmagid and H. S. Dhillon are with Wireless@VT, Department of ECE, Virginia Tech, Blacksburg, VA. Email: \{maelaziz,\ hdhillon\}@vt.edu. N. Pappas is with the Department of Science and Technology, Link\"{o}ping University, SE-60174 Norrk\"{o}ping, Sweden. Email: nikolaos.pappas@liu.se. The support of the U.S. NSF (Grant CPS-1739642) is gratefully acknowledged. This paper will be presented in part at the IEEE Globecom, 2019 \cite{AbdElmagid2019Globecom_a}. \hfill Manuscript updated: \today.}
%\vspace{-5mm}
}

\maketitle

\begin{abstract}
In this paper, we study a real-time monitoring system in which multiple source nodes are responsible for sending update packets to a common destination node in order to maintain the freshness of information at the destination. Since it may not always be feasible to replace or recharge batteries in all source nodes, we consider that the nodes are powered through \textit{wireless energy transfer} (WET) by the destination. For this system setup, we investigate the optimal online sampling policy (referred to as the \textit{age-optimal policy}) that jointly optimizes WET and scheduling of update packet transmissions with the objective of minimizing the long-term average weighted sum of Age-of-Information (AoI) values for different physical processes (observed by the source nodes) at the destination node, referred to as the \textit{sum-AoI}. To solve this optimization problem, we first model this setup as an average cost Markov decision process (MDP) with finite state and action spaces. Due to the extreme curse of dimensionality in the state space of the formulated MDP, classical reinforcement learning algorithms are no longer applicable to our problem even for reasonable-scale settings. Motivated by this, we propose a deep reinforcement learning (DRL) algorithm that can learn the age-optimal policy in a computationally-efficient manner. We further characterize the structural properties of the age-optimal policy analytically, and demonstrate that it has a threshold-based structure with respect to the AoI values for different processes. We extend our analysis to characterize the structural properties of the policy that maximizes average throughput for our system setup, referred to as the \textit{throughput-optimal policy}. Afterwards, we analytically demonstrate that the structures of the age-optimal and throughput-optimal policies are different. We also numerically demonstrate these structures as well as the impact of system design parameters on the optimal achievable average weighted sum-AoI.
%The age-optimal policy determines whether each slot should be allocated for WET or an update packet transmission from some source node while considering the dynamics of levels of batteries at the source nodes, AoI values, and channel state information (CSI)
\end{abstract}
\section{Introduction} \label{sec:intro}
A typical real-time monitoring system consists of source and destination nodes, where source nodes observe underlying stochastic processes while the destination nodes keep track of the status of these processes through status updates transmitted (often wirelessly) by the source nodes. Examples of the source nodes include Internet of Things (IoT) devices, aggregators and sensors, while of the destination nodes include cellular base stations (BSs) \cite{abd2018role}. The performance of many such real-time systems and applications depends upon how {\em fresh} the status updates are when they reach the destination nodes. In practice, the timely delivery of the measurements to the destination nodes is greatly restricted by the limited energy budget of the source nodes and the pathloss of the wireless channel between the source and destination nodes. Specifically, this could result in the loss or out-of-order reception of the measurements at the destination nodes. Consequently, the staleness of information status at the destination nodes increases, which eventually degrades the performance of such real-time applications. 

Since it is highly inefficient or even impractical to replace or recharge batteries in many source nodes, energy harvesting solutions have been considered to enable a self-perpetuating operation of communication networks by supplementing or even circumventing the use of replaceable batteries in the source nodes. Due to its ubiquity and cost efficient implementation, radio-frequency (RF) energy harvesting has quickly emerged as an appealing solution for charging low-power source nodes (especially the ones that are deployed at difficult-to-reach places where other sources of energy harvesting may not be available). This necessitates designing efficient transmission policies for freshness-aware RF-powered communication systems, which is the main objective of this paper. Towards this objective, we use the concept of AoI to quantify the freshness of information at the destination nodes \cite{kaul2012real}. This raises the obvious question of optimally scheduling packet transmissions from these RF-powered source nodes with the objective of minimizing the average AoI at the destination nodes, subject to the energy causality constraints at the source nodes. To address this question, we provide a novel reinforcement learning framework in which we: i) propose a computationally-efficient approach to characterize the age-optimal transmission policy numerically, ii) analytically derive the structural properties of the age-optimal policy, and iii) analytically characterize key differences in the structural properties of the age-optimal and throughout-optimal policies.
%The performance of several real-time applications is mainly driven by how fresh the collected data measurements of the source nodes (e.g., Internet of Things (IoT) devices, aggregators or sensors) are when they reach the destination nodes (e.g., cellular base stations (BSs)) \cite{abd2018role}. The timely delivery of the measurements to the destination nodes is greatly restricted by the limited-energy resources available at source nodes, which also reduce the life time of these nodes. Specifically, due to the energy-constrained nature of the source nodes, several losses in measurement transmissions or out-of-order receptions at the destination nodes occur.
%
%This raises the following open question: {\it How can the status update packet transmissions from different RF-powered source nodes be scheduled with the objective of minimizing the average AoI at the destination nodes, subject to the energy causality constraints at the source nodes?}.
\subsection{Related Work} 
First introduced in \cite{kaul2012real}, AoI is a new metric that quantifies the freshness of information at a destination node due to the transmission of update packets by the source node. Formally, AoI is defined as the time passed since the latest successfully received update packet at the destination was generated at the source node. Under a simple queue-theoretic model in which randomly generated packets arrive at the source according to a Poisson process and then are transmitted to the destination using a first-come-first-served (FCFS) discipline, the authors of \cite{kaul2012real} characterized the average AoI expression. Afterwards, a series of works \cite{yates2012real,kam2013age,Modiano2015
,costa2016age,
chen2016age,Basel,kosta2017age,
javani2019age} aimed at characterizing the average AoI and its variations (e.g., Peak Age-of-Information (PAoI) \cite{costa2016age,
chen2016age,Basel} and Value of Information of Update (VoIU) \cite{kosta2017age}) for adaptations of the queueing model studied in \cite{kaul2012real}. Another direction of research \cite{sun2017update,ABedewy2016,kadota2016,
chen2019benefits,
talak2017,8025774,abd2018average,AoI_UAV,
AbdElmagid2019Globecom_b,Yifan,
zhou2018joint,abdel2018ultra,Buyukates_ulu,Zhong,Maatouk
,bastopcu_ulu,Bo_Ji,Hasan,
Tasmeen,Elif_reinforce} focused on employing AoI as
a performance metric for different communication systems that deal with time critical information while having
limited resources, e.g., multi-server information-update systems \cite{ABedewy2016}, broadcast networks \cite{kadota2016,
chen2019benefits}, multi-hop networks \cite{talak2017}, cognitive networks \cite{8025774}, unmanned aerial vehicle (UAV)-assisted communication systems \cite{abd2018average,AoI_UAV,
AbdElmagid2019Globecom_b}, IoT networks \cite{abd2018role,Yifan,
zhou2018joint}, ultra-reliable low-latency vehicular networks \cite{abdel2018ultra}, and multicast networks \cite{Buyukates_ulu}. Particularly, the objective of this research direction was to characterize optimal
policies that minimize average AoI, referred to as \textit{age-optimal polices}, by applying different tools from optimization theory.
%\cite{biason2017battery,Abd-Elmagid2016,li2019online,salehi2019finite}
%how frequently the information status at a destination node is updated due to update packet transmissions from a source node.

Different from \cite{sun2017update,ABedewy2016,kadota2016,
chen2019benefits,
talak2017,8025774,abd2018average,AoI_UAV,
AbdElmagid2019Globecom_b,Yifan,
zhou2018joint,abdel2018ultra,Buyukates_ulu,Zhong,Maatouk
,bastopcu_ulu,Bo_Ji,Hasan,
Tasmeen,Elif_reinforce}, another line of research \cite{yates2015lazy,2_3,10,2_2,2_1,2,3,8,shahab_2,
George2019,chen2019age,2_5,2_8} focused on the class of problems in which the source node is powered by energy harvesting under various system settings. The objective of this line of research was to investigate age-optimal offline/online policies for update packet transmissions subject to the energy causality constraint at the source under various assumptions regarding the battery size, transmission time of update packets and channel modeling. Specifically, the infinite battery capacity case was studied in \cite{yates2015lazy,10,2_3,2_2,chen2019age} whereas \cite{2_1,2,3,8,shahab_2,George2019} considered the case of finite battery capacity. Different from \cite{10,2_2,2_1,2,3,8} where it was assumed that each update packet could be transmitted to the destination instantly subject to the energy causality constraint, \cite{yates2015lazy,George2019,chen2019age} considered stochastic transmission time and \cite{2_3} studied the non-zero fixed transmission time case. While \cite{yates2015lazy,2_3,10,2_1,2,3,8,shahab_2} considered error-free channel models, i.e., every update packet transmission is successfully received at the destination, a noisy channel model was considered in \cite{2_2,George2019,chen2019age}. A common model of the energy harvesting process in \cite{yates2015lazy,2_3,10,2_2,2_1,2,3,8,shahab_2,
George2019,chen2019age} is an external point process (e.g., Poisson process) independent from all the system design parameters. In contrast, when the source node is powered by RF energy harvesting, as considered in this paper, the energy harvested at the source is a function of the temporal variation of the channel state information (CSI). This, in turn, means that the age-optimal polices studied in \cite{yates2015lazy,2_3,10,2_2,2_1,2,3,8,shahab_2,
George2019,chen2019age} are not directly applicable to this setting. In particular, one needs to incorporate CSI statistics in the process of decision-making, which adds another layer of complexity to the analysis of age-optimal policies for such settings.

Before going into more details about our contributions, it is instructive to note that the problem of age-optimal policy in wireless powered communications systems has been studied very recently in \cite{2_5,2_8} for a single source-destination pair model. Specifically, assuming that the WET and update packet transmissions are performed in orthogonal channels and a dedicated energy source broadcasts RF signals continuously over time to charge the source node, \cite{2_5} proposed a greedy policy in which the source node transmits an update packet using all its available energy (i.e., without any energy management) only if its battery is fully charged. On the other hand, considering that the source node is equipped with infinite battery capacity, \cite{2_8} investigated the optimal transmission policy that minimizes average AoI using tools from convex optimization. The optimization problem in \cite{2_8} was subject to a constraint which guarantees that the long-term average harvested energy is greater than the energy required for update packet transmissions with some probability. Clearly, neither of the policies proposed in \cite{2_5,2_8} took into account the evolution of the battery level at the source and the variation of CSI over time in the process of decision-making. Different from these, we consider a more general model in which multiple RF-powered source nodes are deployed to potentially sense different physical processes. For this setting, this paper makes the first attempt to: 1) characterize the online age-optimal sampling policy while considering the dynamics of batteries, AoI values for different processes and CSI, and 2) analytically characterize key differences between the structures of the online age-optimal and throughput-optimal polices. More details on the contributions in this paper are provided next.
%Clearly, the greedy transmission policy of \cite{2_5} is energy-inefficient (due to the continuity of broadcasting RF energy by the dedicated energy source regardless of the amount of energy available at the source), and the resulting average AoI from the proposed policy of \cite{2_8} is achieved probabilistically (which greatly degrades the performance of the associated monitoring systems especially those are related to human safety applications). 
\subsection{Contributions}
This paper studies a real-time monitoring system in which multiple source nodes are supposed to keep the status of their observed physical processes fresh at a common destination node by transmitting update packets frequently over time. Furthermore, each source node is assumed to be powered by harvesting energy from RF signals broadcast by the destination node. For this setup, our main contributions are listed next.
%This paper studies a real-time monitoring system in which multiple source nodes are supposed to keep the information status at a destination node about their observed physical processes as fresh as possible through sending update packets frequently over time.

{\it A novel DRL algorithm for optimizing average weighted sum-AoI.} Given an importance weight for each physical process at the destination node, we study the long-term average weighted sum-AoI (i.e., sum of AoI values for different processes at the destination node) minimization problem in which WET and scheduling of update packet transmissions from different source nodes are jointly optimized. To tackle this problem, we model it as an average cost MDP with finite state and action spaces. In particular, the MDP determines whether each time slot should be allocated for WET or an update packet transmission from one of the source nodes. This decision is based on the available energies at the source nodes (or their {\em battery levels}), the AoI values of different processes at the destination node, and the CSI. Due to the extreme curse of dimensionality in the state space of the formulated MDP, it is computationally infeasible to characterize the age-optimal policy using classical reinforcement learning algorithms \cite{bertsekas2011dynamic} such as relative value iteration algorithm (RVIA), value iteration algorithm (VIA) or policy iteration algorithm (PIA). To overcome this hurdle, we propose a novel DRL algorithm that can learn the age-optimal policy in a computationally-efficient manner.  

{\it Analytical characterization for the structural properties of the age-optimal policy.} By analytically establishing the monotonicity property of the value function associated with the formulated MDP, we show that the age-optimal policy is a threshold-based policy with respect to each of the AoI values for different processes. Moreover, for the single source-destination pair model (i.e., the case of having a single source node), our results demonstrate that the age-optimal policy is a threshold-based policy with respect to each of the system state variables, i.e., the battery level at the source, the AoI at the destination and the channel power gains. This result is of interest on its own because of the relevance of the source-destination pair model in plethora of applications, such as predicting and controlling forest fires, safety of an intelligent transportation system, and efficient energy utilization in future smart homes. Not surprisingly, this model has been of interest in a large proportion of the prior work on AoI. Furthermore, this result allows us to analytically demonstrate the key differences between the structures of the age-optimal and throughput optimal policies.  
%This result on its own is of significant interest for many applications due to the fact that the single source-destination pair model may be sufficient to enable a diverse set of applications, e.g., predict and control forest fires, safety of an intelligent transportation system, and efficient energy consumption in future smart homes.

{\it System design insights.} Our results provide several useful system design insights. For instance, they show that the differences between the structures of the age-optimal and throughput-optimal policies in the single source-destination pair model mainly depend upon the AoI value of the observed process at the destination node. In particular, while the age-optimal and throughput-optimal policies have different structures when the AoI value is large, these differences start to vanish as the AoI value decreases. After showing the convergence of our proposed DRL algorithm, our numerical results also demonstrate the impact of system design parameters, such as the capacity of batteries and the size of update packets, on the achievable average weighted sum-AoI. Specifically, they reveal that the achievable average weighted sum-AoI by the DRL algorithm is  monotonically decreasing (monotonically increasing) with the capacity of batteries (the size of update packets).
%In particular, when the AoI value is small, the age-optimal and throughput-optimal policies have a fairly-close structures. On the other hand, the age-optimal policy has a completely different structure from the throughput-optimal one for large values of AoI.
\subsection{Organization}
The rest of the paper is organized as follows. Section \ref{sec:Model} presents our system model. The long-term weighted sum-AoI minimization problem is then formulated in Section \ref{sec: age_form}, where a DRL algorithm is proposed to obtain its solution. Afterwards, we present our analysis used to characterize the structural properties of the age-optimal policy in Section \ref{sec:analyt}. Using the analytical results derived in Section \ref{sec:analyt}, the key differences between the structural properties of the age-optimal and throughput-optimal policies in the single source-destination pair model are demonstrated in Section \ref{sec: comparison}. Section \ref{sec:results} verifies our analytical findings from Sections \ref{sec:analyt} and \ref{sec: comparison} as well as evaluates the performance of our proposed DRL algorithm numerically. Finally, Section \ref{sec:con} concludes the paper.
\section{System Model}\label{sec:Model}
\subsection{Network Model}
We study a real-time monitoring system in which a set $\mathcal{I}$ of $N$ source nodes is deployed to observe potentially different physical processes, such as temperature or humidity. Each source node is supposed to keep the information status of its observed process at a destination node (for instance, a cellular BS) fresh by sending status update packets over time. In the context of IoT networks, the source node could refer to a single IoT device or an aggregator located near a group of IoT devices, which transmits update packets collected from them to the destination node. The destination node is assumed to have a stable energy source whereas each source node is equipped with an RF energy harvesting circuitry as its only source of energy. In particular, the source nodes harvest energy from the RF signals broadcast by the destination in the downlink such that the energy harvested at source node $i$ is stored in a battery with finite capacity $B_{{\rm max},i}$~joules. The source and destination nodes are assumed to have a single antenna each and operate over the same frequency channel. Hence, at a given time instant, each source node cannot simultaneously harvest wireless energy in downlink and transmit data in uplink.

We consider a discrete time horizon composed of slots of unit length (without loss of generality) where slot $k = 0,1,\ldots$ corresponds to the time duration $[k,k+1)$. Denote by $B_i(k)$ and $A_i(k)$ the amount of available energy at source node $i$ and the AoI of its observed process $i$ at the destination, respectively, at the beginning of time slot $k$. We assume that $A_i(k)$ is upper bounded by a finite value $A_{{\rm max},i}$ which can be chosen to be arbitrarily large, i.e., $A_i(k) \in \{1,2,\cdots,A_{{\rm max},i}\}$. When $A_i(k)$ reaches $A_{{\rm max},i}$, it means that the available information at the destination nodes about process $i$ is too stale to be of any use. In addition, this assumption makes the AoI state space finite, which facilitates the solution of MDP, as will be clarified in the next section. Let $g_i(k)$ and $h_i(k)$ denote the downlink and uplink channel power gains between the destination and source node $i$ over slot $k$, respectively. The downlink and uplink channels are assumed to be affected by quasi-static flat fading, i.e., they remain constant over a time slot but change independently from one slot to another. The locations of the source nodes are known {\em a priori}, and hence their average channel power gains are pre-estimated and known at the destination node. In particular, at the beginning of an arbitrary time slot, the destination node has perfect knowledge about the channel power gains in that slot, and only a statistical knowledge for future slots. This is a very reasonable assumption for many IoT applications.
\subsection{State and Action Spaces}
At the beginning of an arbitrary time slot $k$, the state $s_i(k)$ of a source node $i$ is characterized by its battery level, the AoI of its observed process $i$ at the destination, and its uplink and downlink channel power gains from the destination node, i.e., $s_i(k) \triangleq (B_i(k), A_i(k), g_i(k), h_i(k)) \in \ncalS_i^a$. Note that $\ncalS_i^{a}$ is the state space which contains all the combinations of $B_i(k), A_i(k), g_i(k)\; \text{and}\; h_i(k)$, where the superscript $a$ indicates that it is defined for the average AoI minimization problem. The state of the system at slot $k$ is then given by $s(k) = \left \{s_{i}(k)\right\}_{i \in \mathcal{I}} \in \mathcal{S}^{a}$, where $\mathcal{S}^{a}$ is the system state space. Based on $s(k)$, the action taken at slot $k$ is given by $a(k) \in \ncalA \triangleq \{H, T_1,T_2,\cdots,T_{N}\}$. When $a(k) = H$, slot $k$ is dedicated for WET where the destination broadcasts RF energy signal in the downlink to charge the batteries at the source nodes. Particularly, the amount of energy harvested by an arbitrary source node $i$ can be expressed as 
\begin{align}\label{eq:harv_energy}
E_i^{\rm H}(k) = \eta P g_i(k), 
\end{align}
where $\eta$ is the efficiency of the energy harvesting circuitry and $P$ is the average transmit power by the destination. We assume that $P$ is sufficiently large such that the energy harvested at each source node due to uplink data transmissions by the other source nodes is negligible. On the other hand, when $a(k) = T_i$, slot $k$ is allocated for information transmission where source $i$ sends an update packet about its observed process to the destination. We consider a {\it generate-at-will policy} \cite{sun2017update}, where the source scheduled for transmission generates an update packet at the beginning of the time slot whenever that slot is allocated for information transmission. According to Shannon's formula, when the energy consumed by source $i$ to transmit an update packet of size $S$ in slot $k$ is $E_i^{\rm T}(k)$, its maximum reliable transmission rate is ${\rm log}_2 \left(1 + \frac{h_i(k) E_i^{\rm T}(k)}{\sigma^2} \right)$ bits/Hz (recall that the slot length is unity), where $\sigma^2$ is the noise power at the destination. Hence, the action $T_i$ can only be decided if the battery level at source $i$ satisfies the following condition
\begin{align}\label{eq:trans_energy_cond}
B_i(k)\geq E_i^{\rm T}(k) = \frac{\sigma^2}{h_i(k)}\left(2^{\bar{S}} - 1\right).
\end{align}
%where $\bar{S}=\frac{S}{W}$ and $W$ is the channel bandwidth.
%We consider a {\it generate-at-will policy} \cite{sun2017update}, where whenever a time slot is allocated for information transmission, the source scheduled for transmission generates an update packet at the beginning of that time slot.

In every time slot, the battery level at each source node and the AoI values for different processes at the destination are updated based on the action decided. Specifically, if $a(k) = T_i$, then the battery level at source $i$ decreases by $E_i^{\rm T}(k)$, and the AoI value of its observed process $i$ becomes one (recall that a generate-at-will policy is employed); if $a(k) = H$, then the battery level at source $i$ increases by $E_i^{\rm H}(k)$ and the AoI value of process $i$ increases by one; otherwise, the battery level at source $i$ does not change and the AoI value of process $i$ increases by one. Hence, the evolution of the battery level at source $i$ and the AoI value of its observed process at the destination node can be expressed, respectively, by
\begin{align}\label{eq:batt_evol}
B_i(k + 1) = \begin{cases}
\begin{aligned}
&B_i(k) - E_i^{\rm T}(k),\; &&\text{if}\; a(k) = T_i,\\
&{\rm min} \left \{B_{{\rm max},i}, B_i(k)+E_i^{\rm H}(k) \right \},\; &&\text{if}\; a(k) = H,\\
&B_i(k),\; &&\text{otherwise}.
\end{aligned}
\end{cases}
\end{align}
\begin{align}\label{eq:AoI_evol}
A_i(k + 1) = \begin{cases}
\begin{aligned}
&1,\; &&\text{if}\; a(k) = T_i,\\
&{\rm min}\left \{A_{{\rm max},i}, A_i(k)+1 \right \},\; &&\text{otherwise}.
\end{aligned}
\end{cases}
\end{align}

\begin{figure}[t!]
\centering
\includegraphics[width=0.5\textwidth]{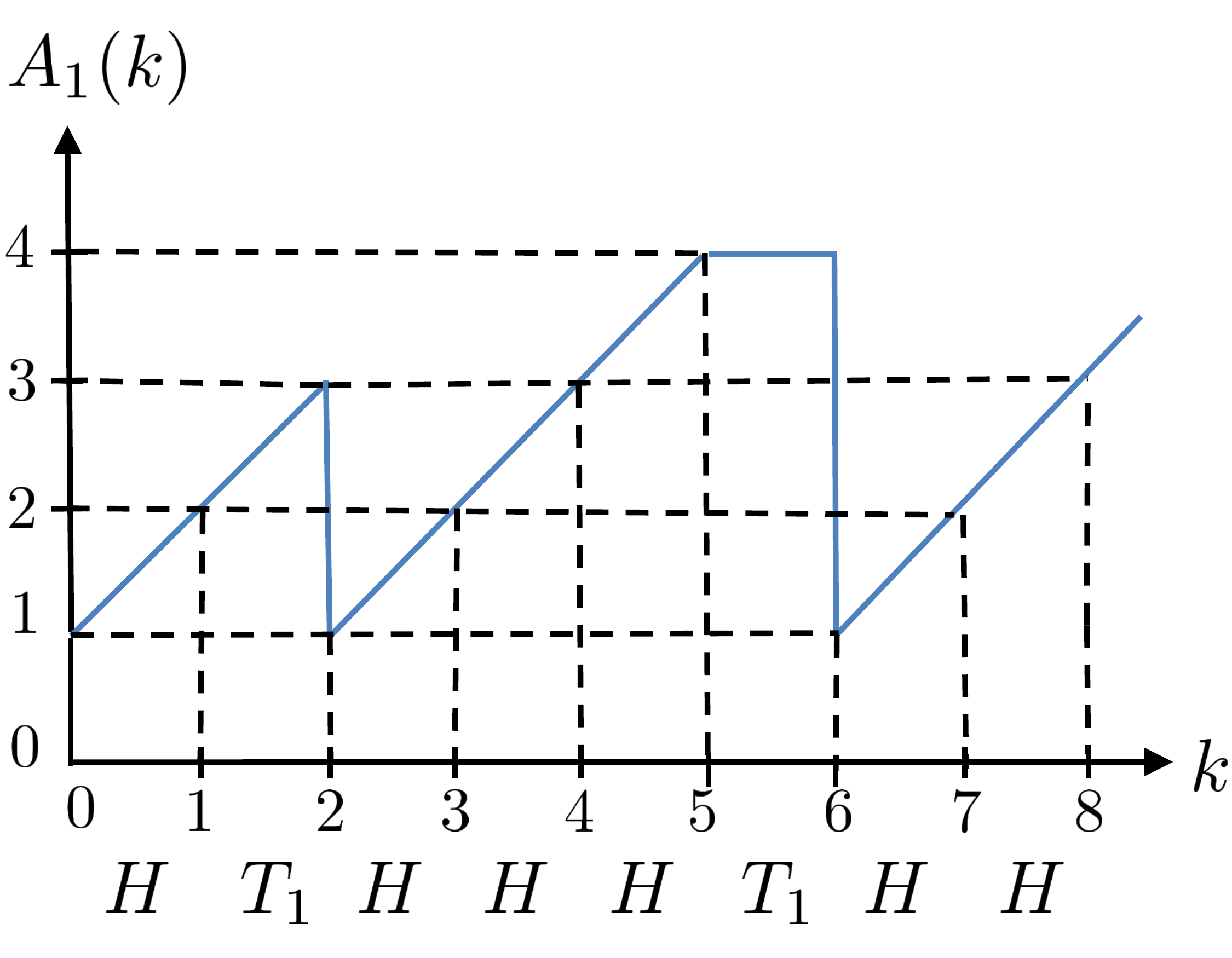}
\caption{AoI evolution vs. time when $N = 1$ and $A_{{\rm max},1} = 4$.}
\label{f:AoI_illustration}
\end{figure}

To help visualize (\ref{eq:AoI_evol}), Fig. \ref{f:AoI_illustration} shows the AoI evolution for process $1$ as a function of actions taken over time when $N = 1$ and $A_{{\rm max},1} = 4$.
\section{Problem Formulation and Proposed Solution}\label{sec: age_form}
\subsection{Problem Statement}
Our objective is to obtain the optimal policy, which specifies the actions taken at different states of the system over time, achieving the minimum average weighted sum-AoI, i.e., sum of AoI values for different processes at the destination. Particularly, a policy $\pi = \{\pi_0,\pi_1,\cdots\}$ is a sequence of probability measures of actions over the state space. For instance, the probability measure $\pi_k$ specifies the probability of taking action $a(k)$, conditioned on the sequence $s^k$ which includes the past states and actions, and the current state, i.e., $s^k \triangleq \left \{s(0),a(0),\cdots,s(k - 1),a(k - 1),s(k) \right \}$. Formally, $\pi_k$ specifies $\P(a(k)\g s^k)$ such that $\sum_{a(k) \in \ncalA(s(k))}{\P(a(k)\g s^k)} = 1$, where $\mathcal{A}(s(k))$ is the set of possible actions at state $s(k) \in \ncalS^a$. The policy $\pi$ is said to be stationary when $\P(a(k)\g s^k) = \P\left(a\left(k\right)\g s\left(k\right)\right), \forall k$, and is called deterministic if $\P(a(k)\g s^k) = 1$ for some $a(k) \in \ncalA(s(k))$. Under a policy $\pi$, the long-term average AoI of process $i$ at the destination starting from an initial state $s(0)$ can be expressed as 
\begin{align}\label{average_AoI}
\bar{A}_i^{\pi} \triangleq \limsup_{K\to \infty} \frac{1}{K+1} \sum_{k = 0}^K \mathbb{E}\left[A_i(k) \g s(0)\right], 
\end{align}
where the expectation is taken with respect to the channel conditions and the policy. Our goal is to find the optimal policy $\pi^\star$, referred to as the \textit{age-optimal policy}, that minimizes the average weighted sum-AoI such that
\begin{align}\label{Optim_prob}
\pi^\star = {\rm arg}\; \underset{\pi}{\rm min} \sum\limits_{i \in \mathcal{I}} {\theta_i \bar{A}_i^{\pi}},
\end{align}
where $\theta_i \geq 0$ and $\sum_{i = 1}^{N}{\theta_i} = 1$. Here, $\theta_i$ is a weight accounting for the importance of process $i$ at the destination node. For instance, if $\theta_i = 1$, then only source node $i$ is considered in the optimization problem. Clearly, the optimal strategy $\pi^*$ is to select whether each time slot is dedicated for WET $(a = H)$ or is allocated for an update packet transmission from source $i$ $(a = T_i)$, depending upon the AoI value of process $i$, and the battery level and channel power gains at source $i$. Hence in this scenario, the achievable average AoI values for other processes are given by $\bar{A}_{j}^{\pi^{\star}} = A_{{\rm max},j}, \forall j \neq i$.
\subsection{MDP Formulation}
Due to the nature of evolution of the battery level at source $i$ and the AoI value of process $i$ at the destination (as described by (\ref{eq:batt_evol}) and (\ref{eq:AoI_evol}), $\forall i \in \mathcal{I}$), and the independence of channel power gains over time, the problem can be modeled as an MDP. In particular, we denote by $b_i(k) \in \{0,1,\cdots,b_{{\rm max},i}\}$ the discrete battery level at source $i$ at the beginning of slot $k$, where $b_{{\rm max},i}$ represents the maximum amount of energy quanta that can be stored in the battery at source $i$ such that each energy quantum contains $\frac{B_{{\rm max},i}}{b_{{\rm max},i}}$ joules. In this case, the quantities $E_i^{\rm T}(k)$ and $E_i^{\rm H}(k)$ in (\ref{eq:batt_evol}) should be replaced by two integer variables expressed in terms of energy quanta. Therefore, by defining $e_i^{\rm T}(k) \triangleq \left \lceil \frac{b_{{\rm max},i}}{B_{{\rm max},i}} E_i^{\rm T}(k)\right \rceil$ and $e_i^{\rm H}(k) \triangleq \left \lfloor \frac{b_{{\rm max},i}}{B_{{\rm max},i}} E_i^{\rm H}(k)\right \rfloor$, the dynamics of the battery at source $i$ for the discrete model can be expressed as
\begin{align}\label{eq:batt_evol_dis}
b_i(k + 1) = \begin{cases}
\begin{aligned}
&b_i(k) - e_i^{\rm T}(k),\; &&\text{if}\; a(k) = T_i,\\
&{\rm min}\left \{b_{{\rm max},i}, b_i(k)+e_i^{\rm H}(k) \right \},\; &&\text{if}\; a(k) = H,\\
&b_i(k),\; &&\text{otherwise},
\end{aligned}
\end{cases}
\end{align}
where we used the ceiling and floor in the definitions of $e_i^{\rm T}(k)$ and $e_i^{\rm H}(k)$ to obtain a lower bound to the performance of the continuous system. Clearly, an upper bound to the performance of the continuous system can be obtained by reversing the use of the floor and ceiling in the definitions of $e_i^{\rm T}(k)$ and $e_i^{\rm H}(k)$. Similarly, if the channel power gains are modeled by continuous random variables, we divide their support into a finite number of intervals with the same probability according to the fading probability density function (PDF). In this sense, the problem is modeled as a \textit{finite-state finite-action MDP} with state $s(k) \triangleq \left \{(b(k), A(k), g(k), h(k))\right\}_{i \in \mathcal{I}} \in \ncalS_{\rm d}^a$ (the state space of the discrete model) and action $a(k) \in \mathcal{A}(s(k)) \subseteq \mathcal{A}$. Since there exists an optimal stationary deterministic policy for solving finite-state finite-action MDPs \cite{bertsekas2011dynamic}, we aim at investigating that age-optimal stationary deterministic policy in the sequel and omit the time index. Note that as the number of discrete levels for both batteries and channel power gains increase, the discrete model can be considered as a good approximation for the continuous one, but this comes at the expense of a high computational complexity to characterize $\pi^\star$.

Due to taking an action $a$, the transition probability of moving from state $s_i = (b_i,A_i,g_i,h_i)$ to state $s'_i=(b'_i,A'_i,g'_i,h'_i)$ at source node $i$ is given by
\begin{align}\label{transprob}
\nonumber \P \left(s'_i \g s_i, a \right)  &\triangleq \P \left(b'_i, A'_i, g'_i, h'_i \g b_i, A_i, g_i, h_i, a \right)\\
\nonumber &\a \P \left(b'_i, A'_i \g b_i, A_i, g_i, h_i, a \right) \P(g'_i) \P(h'_i)\\
&\b \P \left(b'_i \g b_i, g_i, h_i, a\right) \P \left(A'_i \g A_i, a \right) \P(g'_i) \P(h'_i),
\end{align}
%where $\pi(s)$ denotes the action taken at state $s$ according to $\pi$. 
where step (a) follows from the independence of the channel power gains over time and from other random variables, where $\P(g'_i)$ and $\P(h'_i)$ denote the probability mass functions for the downlink and uplink channel power gains (after discretization if they were expressed originally by continuous random variables), respectively. Step (b) follows since given $s_i$ and $a$, the next battery level $b'_i$ and the value of AoI $A'_i$ can be obtained deterministically, separately from each other. Specifically, $b'_i$ only depends on the current battery level and channel power gains, i.e., $(b_i,g_i,h_i)$, and $A'_i$ only depends upon its current value $A_i$. Thus, from (\ref{eq:AoI_evol}) and (\ref{eq:batt_evol_dis}), $b'_i$ and $A'_i$ can be determined, respectively, as
\begin{align}\label{eq:batt_transprob}
\P(b'_i \g b_i, g_i, h_i, a) = \begin{cases}
\begin{aligned}
& \mathbbm{1} \left(b'_i = b_i - e_i^{\rm T}\right),\; \text{if}\; a = T_i,\\
& \mathbbm{1}\left(b'_i = {\rm min} \left \{b_{{\rm max},i}, b_i + e_i^{\rm H} \right \}\right),\;\text{if}\; a = H,\\
& \mathbbm{1}\left(b'_i = b_i\right),\;\text{otherwise},
\end{aligned}
\end{cases}
\end{align}
\begin{align}\label{eq:AoI_transprob}
\P(A'_i \g A_i, a) = \begin{cases}
\begin{aligned}
&\mathbbm{1} \left(A'_i = 1 \right),\; \text{if}\; a = T_i,\\
&\mathbbm{1} \left(A'_i = {\rm min}\left \{A_{{\rm max},i}, A_i+1 \right\}\right),\; \text{otherwise},
\end{aligned}
\end{cases}
\end{align}
where $\mathbbm{1}(\cdot)$ is the indicator function. The overall transition probability of moving from state $s = \{ s_i\}_{i \in \mathcal{I}}$ to state $s' = \{ s'_i\}_{i \in \mathcal{I}}$, after taking an action $a$, can then be expressed as
\begin{align}\label{overalltransprob}
\P \left(s' \g s, a \right) \a \prod_{i \in \mathcal{I}}{\P \left(s'_i \g s_i, a \right)},
\end{align}
where (a) follows from the fact that given action $a$, the state of each source node evolves separately from the other source nodes. The following Lemma characterizes the optimal policy $\pi^{\star}$ satisfying (\ref{Optim_prob}).
\begin{lemma}\label{Lemma:1}
 The optimal policy $\pi^\star$ can be evaluated by solving the following Bellman's equations for average cost MDPs \cite{bertsekas2011dynamic}:
\begin{align}\label{belman_equation}
\bar{A}^\star + V(s) = \underset{a \in \mathcal{A}(s)}{\rm min} Q(s,a), s \in \mathcal{S}_{\rm d}^a,
\end{align}
where $\bar{A}^\star$ is the achievable optimal average AoI under $\pi^\star$ which is independent of the initial state $s(0)$, $V(s)$ is the value function, and $Q(s,a)$ is the $Q$-function $\big($also referred to as the $Q$-factors, $\forall s \in \mathcal{S}_{\rm d}^a$ and $a \in \mathcal{A}(s)\big)$, which is the expected cost resulting from taking action $a$ in state $s$, i.e.,
\begin{align}\label{Q_func}
Q(s,a) = \sum\limits_{i \in \mathcal{I}}{\theta_i A_i} + \sum\limits_{s' \in \mathcal{S}_{\rm d}^a} {\P(s' \g s, a) V(s')},
\end{align}
where $\P(s' \g s, a)$ is evaluated using $(\ref{overalltransprob})$. In addition, the optimal action taken at state $s$ is given by
\begin{align}\label{eq:optimal_policy}
\pi^\star(s) = {\rm arg} \underset{a \in \mathcal{A}(s)}{\rm min} Q(s,a).
\end{align}
\end{lemma}

Note that the weak accessibility condition holds for our problem, and hence a solution for the Bellman's equations in Lemma \ref{Lemma:1} is guaranteed to exist \cite{bertsekas2011dynamic}. Characterizing the optimal policy by solving Bellman's equations using classical reinforcement learning algorithms \cite{bertsekas2011dynamic} (e.g., VIA, PIA or RVIA) requires to evaluate the policy improvement setup in (\ref{eq:optimal_policy}) for each state at each iteration. Defining $G_i$ and $H_i$ as the number of discrete values that the state variables $g_i$ and $h_i$ can take, respectively, the number of states inside the state space $\mathcal{S}^a$ can then be computed as $|\mathcal{S}^a| = \underset{i \in \mathcal{I}}{\prod} \left(A_{{\rm max},i} G_i H_i\left(b_{{\rm max},i} + 1\right)\right)$. Clearly, for a reasonable number of both the discrete values for each state variable (i.e., $A_{{\rm max},i}, G_{i}, H_i,$ and $b_{{\rm max},i} + 1$) and the source nodes deployed in the network $(N)$, the state space will have a massive number of states. For instance, if we consider that each state variable can only take $10$ values and there are three source nodes in the network, then the number of states becomes $10^{12}$. As a result, it becomes computationally infeasible to obtain the optimal policy using classical reinforcement learning algorithms as the number of states increases (due to either increasing the number of discrete values for each state variable or the number of source nodes). This calls for investigating new approaches for characterizing the optimal policy in such large-scale setups. In order to overcome this problem, we propose a DRL algorithm to obtain the age-optimal policy numerically in the next subsection. We will also derive several key structural properties of the age-optimal policy analytically in Section \ref{sec:analyt}.
% In order to obtain the age-optimal policy by applying standard optimization techniques such that the Value Iteration Algorithm or the Policy Iteration Algorithm \cite{bertsekas2011dynamic}, we discretize the battery level and the channel power gains.
\subsection{Deep Reinforcement Learning for Optimizing AoI}
DRL is suitable for our problem
since it can reduce the dimensionality of the large state space while learning the optimal policy at the same time \cite{mnih2015human}. The proposed DRL algorithm has two components: i) an artificial neural network (ANN), that reduces the dimension of the system state space by extracting its useful features, and ii) a reinforcement component, which is used to find the best policy based on the ANN's extracted features.
%as shown in Fig [].
%as shown in Fig. \ref{fig:deepRL}.
%\begin{figure}[t!]
%    \centering
%    \includegraphics[width=0.75\columnwidth]{Figures/DeepRL2}
%    \caption{\small The deep RL architecture.}
%    \label{fig:deepRL}
%\end{figure}

The reinforcement learning component is represented by the $Q$-learning algorithm \cite{bertsekas2011dynamic}. As per the $Q$-learning algorithm, an update step for the $Q$-function value of the current state is performed at the beginning of each time slot, based on the action taken as well as the resulting next state. In particular, at the beginning of slot $k + 1$, the update step of the $Q$-learning algorithm for our average cost MDP can be expressed as \cite{bertsekas2011dynamic}:
\begin{align}\label{eq:Bellman}
& Q_{k+1}\left(s(k),a(k)\right) = Q_{k}\left(s(k),a(k)\right) + \nonumber \\ & \alpha(k) \left(c(k) + \min_{\bar{a} \in \mathcal{A}\left(s\left(k+1\right)\right)} Q_{k}\left(s\left(k+1\right),\bar{a}\right) - \min_{\bar{a} \in \mathcal{A}\left(\bar{s}\right)} Q_{k}\left(\bar{s},\bar{a}\right) - Q_{k}\left(s(k),a(k)\right) \right),
\end{align}
where $c(k) = \sum\limits_{i \in \mathcal{I}}{\theta_i A_i(k)}$ represents the resulting cost from taking  action $a(k)$ in state $s(k)$ at slot $k$, $\alpha(k)$ is the learning rate at slot $k$, and $\bar{s}$ is the special state, which remains fixed over all the iterations and can be chosen arbitrarily. Note that \eqref{eq:Bellman} results from applying the $Q$-learning method to the relative value iteration of the $Q$-factors for average cost MDPs \cite{bertsekas2011dynamic}. The sequence of values $\underset{\bar{a} \in \mathcal{A}\left(\bar{s}\right)}{\rm min}Q_{k}\left(\bar{s},\bar{a}\right)$ is expected to converge to the optimal average AoI $\bar{A}^{\star}$ under the following conditions \cite{powell2007approximate}: i) $\sum_{k = 1}^{\infty}{\alpha(k)}$ is infinite and $\sum_{k = 1}^{\infty}{(\alpha(k))^2}$ is finite , ii) all potential state-action pairs are visited infinitely often, and iii) the state transition probability is stationary under the optimal stationary policy. By applying the update step in \eqref{eq:Bellman}, the system can always {\it exploit} the learning process by taking the action which minimizes the long-term average cost, i.e., the action that minimizes the $Q$-function value of the current state. On the other hand, according to condition ii), the system has to {\it explore} all state-action pairs for the convergence of the algorithm. Thus, an $\epsilon$-greedy policy has to be employed \cite{mnih2015human}, where a random action is decided at the current state with probability $0<\epsilon <1$ with the objective of exploring the environment rather than exploiting the learning process. Meanwhile, the value of $\epsilon$ could be reduced to $0$ as the learning goes in order to ensure that the learning process is exploited efficiently, i.e., not too much time is spent on exploring the environment.

Using the $Q$-learning algorithm (presented above) alone to characterize the optimal policy is efficient for cases where the system state space has a relatively small number of states. However, when the number of states is extremely large (which is the case in our problem), it becomes impractical to store the $Q$-function values for all state-action pairs (a massive memory is required for this) or even ensure that all state action-pairs will be visited so that the convergence can be achieved. Thus, as the cardinality of the (discrete) support set of each state variable and/or the number of source nodes increase in our problem, using $Q$-learning alone to characterize the optimal policy is not sufficient. In order to tackle this hurdle, we employ ANNs which are very effective at extracting features from data points and summarizing them in smaller dimensions. Specifically, a deep $Q$ network approach \cite{mnih2015human} is used in which the learning steps are the same as in $Q$-learning, but the $Q$-function is approximated using an ANN $Q(s,a|\boldsymbol{\beta})$, where $\boldsymbol{\beta}$ is the vector containing the weights of the ANN. We utilize a fully connected (FC) layer, as in \cite{mnih2015human}, to extract abstraction of the state space. In the FC layer, every artificial node of a layer is connected to every artificial node of the next layer via the weight vector $\boldsymbol{\beta}$. The objective is then to find the optimal values for $\boldsymbol{\beta}$ such that the stored $Q$-function by the ANN becomes as close as possible to the optimal $Q$-function. To this end, we define a loss function for any combination of $\left(s(k),a(k),c(k),s(k+1)\right)$, as follows:
\begin{align}
    L(\boldsymbol{\beta}_{k+1}) = \Big(c(k) &+ \min_{\bar{a} \in \mathcal{A}\left(s\left(k+1\right)\right)} Q_{k}\left(s\left(k+1\right),\bar{a}|\boldsymbol{\beta}_{k}
\right) \nonumber \\ &- \min_{\bar{a} \in \mathcal{A}\left(\bar{s}\right)} Q_{k}\left(\bar{s},\bar{a}|\boldsymbol{\beta}_{k}
\right) - Q_{k}\left(s(k),a(k)|\boldsymbol{\beta}_{k+1}
\right)\Big)^2,
\end{align}
\begin{algorithm}[t!]
	\caption{Deep reinforcement learning for average weighted sum-AoI minimization}
	\begin{algorithmic}
	%	[1]\footnotesize 
		\State Initialize a replay memory and an ANN $Q$ with a vector of weights $\boldsymbol{\beta}_0$. 
		\State Observe the initial state $s(0)$ and set $k=0$.
		\State \textbf{Repeat:}
		\State \quad Select an action $ a(k) $:
		\State \quad \quad select a random action $a(k) \in \mathcal{A}(s(k))$ with probability $ \varepsilon $,
		\State \quad \quad otherwise select $ a(k) = {\rm arg}\; \underset{\bar{a}}{\rm min}\; Q\left(s(k),\bar{a}|\boldsymbol{\beta}_k\right) $
		\State \quad Perform action $ a(k) $. 
		\State \quad Evaluate the cost $ c(k) $ and observe the new state $ s(k+1) $.
		\State \quad Store \emph{experience} $ \left\{s(k),a(k),c(k),s(k+1)\right\} $ in the replay memory.
		\State \quad Sample a random batch of experiences $ \left\{\hspace{-0.5mm}\hat{s}(\zeta), \hspace{-0.5mm}\hat{a}(\zeta),\hspace{-0.5mm}\hat{c}(\zeta),\hspace{-0.5mm}\hat{s}(\zeta+1)\hspace{-0.5mm}\right\} $ from the replay memory.
		\State \quad Calculate the set of target values $\{t(\zeta)\}$ corresponding to the experiences of the sampled 
		\State \quad batch:
		\State \quad \quad $ t(\zeta)= \hat{c}(\zeta) + \underset{\bar{a} \in \mathcal{A}\left(\hat{s}\left(\zeta+1\right)\right)}{\rm min}Q\left(\hat{s}\left(\zeta+1\right),\bar{a}|\boldsymbol{\beta}_{k}
\right) - \underset{\bar{a} \in \mathcal{A}\left(\bar{s}\right)}{\rm min}Q\left(\bar{s},\bar{a}|\boldsymbol{\beta}_{k}
\right).$
		\State \quad Train the network $ Q $ using the gradient in \eqref{eq:gradient}.
		\State \quad $ k = k+1 $.
		\State \textbf{Until} convergence to some value of average weighted sum-AoI.
	\end{algorithmic}
	\label{Algorithm:DeepRL}
\end{algorithm}
where subscript $k+1$ is the time slot at which the weights are updated.
Furthermore, a \emph{replay memory} is used to save the evaluation of the state, action, and cost of past \emph{experiences}, i.e., past state-action pairs and their resulting costs. In particular, after every time slot, we sample a random batch of a finite number of past experiences from the replay memory, and the gradient of the ANN's weights using this batch is evaluated as follows:
\begin{align}\label{eq:gradient}
\nabla_{\boldsymbol{\beta}_{k+1}} L(\boldsymbol{\beta}_{k+1}) &= \Big(c(k) + \min_{\bar{a} \in \mathcal{A}\left(s\left(k+1\right)\right)} Q_{k}\left(s\left(k+1\right),\bar{a}|\boldsymbol{\beta}_{k}
\right) \nonumber \\ &- \min_{\bar{a} \in \mathcal{A}\left(\bar{s}\right)} Q_{k}\left(\bar{s},\bar{a}|\boldsymbol{\beta}_{k}
\right) - Q_{k}\left(s(k),a(k)|\boldsymbol{\beta}_{k+1}
\right) \Big)\times\nabla_{\boldsymbol{\beta}_{k+1}}Q_k(s(k),a(k)|\boldsymbol{\beta}_{k+1}).
\end{align}
The weights of the ANN are then trained using this loss function. Note that it has been shown in \cite{mnih2015human} that using the batch method and replay memory improves the convergence of DRL. Algorithm \ref{Algorithm:DeepRL} summarizes the steps of the proposed DRL algorithm.
%and Fig. [] shows the architecture of the deep reinforcement learning algorithm.

So far, we have presented our proposed approach to obtain the optimal policy numerically. In the next section, we explore the structural properties of the age-optimal policy $\pi^\star$ analytically. 
\section{Structural Properties of the Age-optimal Policy}\label{sec:analyt}
In this section, we derive the structural properties of the age-optimal policy $\pi^\star$ analytically using the VIA. Note that the obtained analytical results can be derived using the RVIA as well \cite{bertsekas2011dynamic}. For completeness, we start this discussion by summarizing the VIA. According to the VIA, the value function $V(s)$ can be evaluated iteratively such that $V(s)$ at iteration $m$, $m = 1, 2, \cdots$, is computed as 
\begin{align}\label{value_function_itern}
 V(s)^{(m)} = \underset{a \in \mathcal{A}(s)}{\rm min} Q(s,a)^{(m - 1)} = \underset{a \in \mathcal{A}(s)}{\rm min} \left \{\sum\limits_{i \in \mathcal{I}}{\theta_i A_i} + \sum\limits_{s' \in \mathcal{S}^a_{\rm d}} {\P(s' \g s, a) V(s')^{(m - 1)}}\right \},
\end{align}
where $s \in \mathcal{S}_{\rm d}^a$. Hence, the optimal policy at iteration $m$ is given by
\begin{align}\label{policy_itern}
\pi^{\star(m)}(s) = {\rm arg} \underset{a \in \mathcal{A}(s)}{\rm min} Q(s,a)^{(m - 1)}.
\end{align}

As per the VIA, under any initialization of the value function $V(s)^{(0)}$, the sequence $\left \{ V(s)^{(m)}\right\}$ converges to $V(s)$ which satisfies the Bellman's equation in (\ref{belman_equation}), i.e.,
\begin{align}\label{conv}
\underset{m \rightarrow \infty}{\rm lim} V(s)^{(m)} = V(s).
\end{align}

Based on the VIA, the following Lemma characterizes the monotonicity property of the value function with respect to the system state variables. 
\begin{lemma}\label{Lemma:2}
The value function $V(s)$, satisfying the Bellman's equation in (\ref{belman_equation}) and corresponding to the age-optimal policy $\pi^\star$, is non-increasing with respect to the battery level $b_j$, the downlink channel power gain $g_j$ and the uplink channel power gain $h_j, \forall j \in \mathcal{I}$. In contrast, $V(s)$ is non-decreasing with respect to the AoI $A_j, \forall j \in \mathcal{I}$.
\end{lemma}
\begin{IEEEproof}
First, to prove that $V(s)$ is non-increasing with respect to $b_j$, let us define two states $s^{1} = \left \{\left(b_i^1,A_i^1,g_i^1,h_i^1\right) \right \}_{i \in \mathcal{I}}$ and $s^2 = \left \{\left(b_i^2,A_i^2,g_i^2,h_i^2\right) \right \}_{i \in \mathcal{I}}$ where: i) $b_j^1 \leq b_j^2$, ii) $b_i^1 = b_i^2, \forall i \neq j$, and iii) $A_i^1 = A_{i}^2$, $g_i^1 = g_i^2$ and $h_i^1 = h_i^2, \forall i \in \mathcal{I}$. Hence, the objective is to show that $V(s^1) \geq V(s^2)$. According to (\ref{conv}), it is then sufficient to show that $V(s^1)^{(m)} \geq V(s^2)^{(m)}, \forall m$, which we prove using mathematical induction. Particularly, the relation holds by construction for $m = 0$ since it corresponds to the initial values for the value function which can be chosen arbitrarily. Now, we assume that $V(s^1)^{(m)} \geq V(s^2)^{(m)}$ holds for some $m$, and then show that it holds for $V(s^1)^{(m + 1)} \geq V(s^2)^{(m + 1)}$ as well. Particularly, according to (\ref{value_function_itern}) and (\ref{policy_itern}), $V(s^2)^{(m + 1)}$ and  $V(s^1)^{(m + 1)}$ can be expressed, respectively, as
\begin{align}\label{v_s2_m+1}
\nonumber V(s^2)^{(m + 1)} & = \sum\limits_{i \in \mathcal{I}}{\theta_i A_i^2} + \sum\limits_{s^{2'} \in \mathcal{S}^a_{\rm d}} {\P\left(s^{2'} \g s^{2}, \pi^{\star (m)} \left(s^{2}\right)\right) V(s^{2'})^{(m )}} \\
\nonumber & \overset{(a)}{\leq} \sum\limits_{i \in \mathcal{I}}{\theta_i A_i^2} + \sum\limits_{s^{2'} \in \mathcal{S}^a_{\rm d}} {\P\left(s^{2'} \g s^{2}, \pi^{\star (m)} \left(s^{1}\right)\right) V(s^{2'})^{(m )}} \\
& \b \sum\limits_{i \in \mathcal{I}}{\theta_i A_i^2} + C_0 \sum\limits_{g'_1} \sum\limits_{h'_1} \cdots \sum\limits_{g'_N} \sum\limits_{h'_N} { V \left(\left \{b_i^{2'},A'_i,g'_i,h'_i \right \}_{i \in \mathcal{I}}\right)^{(m )}},
\end{align}
\begin{align}\label{v_s1_m+1}
\nonumber V(s^1)^{(m + 1)} &= \sum\limits_{i \in \mathcal{I}}{\theta_i A_i^1} + \sum\limits_{s^{1'} \in \mathcal{S}^a_{\rm d}} {\P\left(s^{1'} \g s^{1}, \pi^{\star (m)} \left(s^{1}\right)\right) V(s^{1'})^{(m )}}\\
&= \sum\limits_{i \in \mathcal{I}}{\theta_i A_i^1} + C_0 \sum\limits_{g'_1} \sum\limits_{h'_1} \cdots \sum\limits_{g'_N} \sum\limits_{h'_N} { V \left(\left \{b_i^{1'},A'_i,g'_i,h'_i \right \}_{i \in \mathcal{I}}\right)^{(m )}},
\end{align} 
where $C_0 = \underset{i \in \mathcal{I}}{\prod}{\P(g'_i) \P(h'_i)}$. Step (a) follows since it is not optimal to take action $\pi^{\star(m)}(s^1)$ in state $s^2$, and step (b) follows from (\ref{transprob})-(\ref{overalltransprob}) where, for a given $\pi^{\star(m)}(s^1)$, the set of values $\{A'_i\}_{i \in \mathcal{I}}$ can be evaluated based on (\ref{eq:AoI_transprob}), and the sets $\{b_i^{2'}\}_{i \in \mathcal{I}}$ and $\{b_i^{1'}\}_{i \in \mathcal{I}}$ are determined using (\ref{eq:batt_transprob}). Note that since $b_i^1 = b_i^2, \forall i \neq j$, we have $b_i^{1'} = b_i^{2'}, \forall i \neq j$. On the other hand since $b_j^{1} \leq b_j^{2}$, we can observe from (\ref{eq:batt_transprob}) that $b_{j}^{1'} \leq b_{j}^{2'}$ for $\pi^{\star(m)}(s_1) \in \mathcal{A}$, and hence $V \left(\left \{b_i^{1'},A'_i,g'_i,h'_i \right \}_{i \in \mathcal{I}}\right)^{(m )} \geq V \left(\left \{b_i^{2'},A'_i,g'_i,h'_i \right \}_{i \in \mathcal{I}}\right)^{(m )}$. Therefore the expression in (\ref{v_s2_m+1}) is less than or equal to $V(s^2)^{(m+1)}$ which implies $V(s^1)^{(m + 1)} \geq V(s^2)^{(m + 1)}$ and indicates that the value function is non-increasing with respect to $b_j$. Note that increasing $g_j$ ($h_j$) increases $e_j^{\rm H}$ (reduces $e_j^{\rm T}$) which leads to a larger amount of energy in the battery at source $j$ at the next time slot and hence a lower value function. This proves that $V(s)$ is non-increasing with respect to $g_j$ and $h_j$, $\forall j \in \mathcal{I}$.

Next, using the same approach, we can show that $V(s)$ is non-decreasing with respect to $A_j$. Now, consider that the two states $s^{1}$ and $s^{2}$ are defined such that: i) $A_j^1 \geq A_j^2$, ii) $A_i^1 = A_i^2, \forall i \neq j$, and iii) $b_i^1 = b_{i}^2$, $g_i^1 = g_i^2$ and $h_i^1 = h_i^2, \forall i \in \mathcal{I}$. The goal is then to show that $V(s^1) \geq V(s^2)$. This can again be proven using mathematical induction by showing that $V(s^1)^{(m)} \geq V(s^2)^{(m)}, \forall m$. In particular, (\ref{v_s2_m+1}) and (\ref{v_s1_m+1}) can be rewritten for this case as
 \begin{align}\label{v_s2_m+1_A}
 V(s^2)^{(m + 1)} \nonumber & \leq \sum\limits_{i \in \mathcal{I}}{\theta_i A_i^2} + \sum\limits_{s^{2'} \in \mathcal{S}^a_{\rm d}} {\P\left(s^{2'} \g s^{2}, \pi^{\star (m)} \left(s^{1}\right)\right) V(s^{2'})^{(m )}} \\
& = \underbrace{\sum\limits_{i \in \mathcal{I}}{\theta_i A_i^2}}_{C_1} + C_0\underbrace{\sum\limits_{g'_1} \sum\limits_{h'_1} \cdots \sum\limits_{g'_N} \sum\limits_{h'_N} { V \left(\left \{b'_i,A_i^{2'},g'_i,h'_i \right \}_{i \in \mathcal{I}}\right)^{(m )}}}_{C_2},
\end{align}
\begin{align}\label{v_s1_m+1_A}
 \nonumber V(s^1)^{(m + 1)} &= \sum\limits_{i \in \mathcal{I}}{\theta_i A_i^1} + \sum\limits_{s^{1'} \in \mathcal{S}^a_{\rm d}} {\P\left(s^{1'} \g s^{1}, \pi^{\star (m)} \left(s^{1}\right)\right) V(s^{1'})^{(m )}}\\
 &= \underbrace{\sum\limits_{i \in \mathcal{I}}{\theta_i A_i^1}}_{C_3} + C_0\underbrace{\sum\limits_{g'_1} \sum\limits_{h'_1} \cdots \sum\limits_{g'_N} \sum\limits_{h'_N} { V \left(\left \{b'_i,A_i^{1'},g'_i,h'_i \right \}_{i \in \mathcal{I}}\right)^{(m )}}}_{C_4},
\end{align} 
where $A_i^{2'} = A_i^{1'}, \forall i \neq j$ due to the fact that $A_i^1 = A_i^2, \forall i \neq j$. Note that we have $C_3 \geq C_1$ by construction since $A_{j}^{1} \geq A_{j}^{2}$. It is then sufficient to show that $C_4 \geq C_2$ for all possible actions $\pi^{\star(m)}(s^1) \in \mathcal{A}(s^{1})$. Specifically, there are two different cases: 1) $\pi^{\star(m)}(s^1) = T_j$, and 2) $\pi^{\star(m)}(s^1) \in \mathcal{A}(s^1) \setminus \{T_j\}$. Based on (\ref{eq:AoI_transprob}), we have $A_j^{1'} = A_j^{2'} = 1$ for the first case and hence $C_4 = C_2$. On the other hand, we have $A_j^{1'} \geq A_j^{2'}$ for the second case, which leads to $C_4 \geq C_2$. Consequently, $V(s^1)^{(m + 1)} \geq V(s^2)^{(m + 1)}, \forall \pi^{\star(m)}(s^1) \in \mathcal{A}(s^1)$ which proves that $V(s)$ is non-decreasing with respect to $A_j, \forall j \in \mathcal{I}$.
\end{IEEEproof}

Based on Lemma \ref{Lemma:2}, the following Theorem characterizes the structure of the age-optimal policy $\pi^\star$ with respect to the AoI values for different processes at the destination node.
\begin{theorem}\label{theorem:1}
Define two states $s^1 = \left \{\left(b_i^1,A_i^1,g_i^1,h_i^1\right) \right \}_{i \in \mathcal{I}}$ and $s^2 = \left \{\left(b_i^2,A_i^2,g_i^2,h_i^2\right) \right \}_{i \in \mathcal{I}}$ such that: i) $A_{j}^{2} \geq A_{j}^{1}$, ii) $A_{i}^{2} = A_{i}^{1}, \forall i \neq j$, and iii) $b_i^1 = b_{i}^2$, $g_i^1 = g_i^2$ and $h_i^1 = h_i^2, \forall i \in \mathcal{I}$. If $\pi^\star(s^1) = T_j$, then $\pi^\star(s^2) = T_j$. 
\end{theorem}
\begin{IEEEproof}
First, we observe that proving $\pi^\star(s^1) = \bar{a}$ implies $\pi^\star(s^2) = \bar{a}$ is equivalent to showing that
\begin{align}\label{struc_prop}
Q(s^2,\bar{a}) - Q(s^2,a') \leq Q(s^1,\bar{a}) - Q(s^1,a'), \forall a' \neq \bar{a}.
\end{align}

This is because if $\bar{a}$ is optimal in state $s^1$, then we have $Q(s^1,\bar{a}) - Q(s^1,a') \leq 0, \forall a' \neq \bar{a}$, which implies $Q(s^2,\bar{a}) \leq Q(s^2,a'),\forall a' \neq \bar{a}$, i.e., taking action $\bar{a}$ is optimal in state $s_2$. Hence, in order to complete the proof, we need to show that (\ref{struc_prop}) holds for all possible choices of $a' \in \mathcal{A}(s^2) \setminus \{T_j\}$ when $\bar{a} = T_j$. To maintain generality, we consider the case where $\mathcal{A}(s^2) = \mathcal{A}$. Particularly, from (\ref{transprob})-(\ref{overalltransprob}) and (\ref{Q_func}), we have
\begin{align}\label{Q_Si,action}
Q(s^n,a) = \sum\limits_{i \in \mathcal{I}}{\theta_i A_i^n} + C_0\underbrace{\sum\limits_{g'_1} \sum\limits_{h'_1} \cdots \sum\limits_{g'_N} \sum\limits_{h'_N} { V \left(\left \{b'_i,A_i^{n'},g'_i,h'_i \right \}_{i \in \mathcal{I}}\right)}}_{C(n,a)}, n \in \{1,2\}.
\end{align} 
 
 According to (\ref{struc_prop}), we first note that the term $\sum\limits_{i \in \mathcal{I}}{\theta_i A_i^n}$ is canceled out from all $Q(s^{n},a)$, $n \in \{1,2\}$ and $a \in \{\bar{a}, a'\}$. When $a = T_j$, we have $A_{j}^{1'} = A_{j}^{2'} = 1$ from (\ref{eq:AoI_transprob}). This means $C(1,T_j)$ will equal $C(2,T_j)$ and (\ref{struc_prop}) will hold if $C(2,a) \geq C(1,a), \forall a \in \mathcal{A} \setminus \{T_j\}$. For any $a \in \mathcal{A} \setminus \{T_j\}$, it follows that $A_{j}^{n'} = {\rm min}\left \{A_{{\rm max},j}, A_j^{n} +1 \right \}$ from (\ref{eq:AoI_transprob}). Since $A_j^{2} \geq A_j^{1}$ from i), we then have $A_{j}^{2'} \geq A_{j}^{1'}$. Now, based on Lemma \ref{Lemma:2} along with taking into account ii) and iii), it follows that $V \left(\left \{b'_i,A_i^{2'},g'_i,h'_i \right \}_{i \in \mathcal{I}}\right) \geq V \left(\left \{b'_i,A_i^{1'},g'_i,h'_i \right \}_{i \in \mathcal{I}}\right)$. Hence, we have $C(2,a) \geq C(1,a)$, which completes the proof.
\end{IEEEproof}
\begin{remark}\label{Remark:1}
For the case of having multiple source nodes deployed in the network, i.e., $N > 1$, Theorem \ref{theorem:1} indicates that the age-optimal policy $\pi^\star$ has a threshold-based structure with respect to each of the AoI state variables for different processes, i.e., $A_j, \forall j \in \mathcal{I}$. For instance, for a fixed combination of state variables excluding $A_{j}$, if $A_{{\rm th},j}$ is the minimum AoI value of process $j$ for which it is optimal to take an action $a = T_j$, then for all states with $A_j \geq A_{{\rm th},j}$, the optimal decision is $T_j$ as well. This is also intuitive, since when the value of AoI for some process becomes large, it is optimal to update the status of information for that process at the destination by sending a new update packet. 
\end{remark}

Note that by checking (\ref{struc_prop}), one can show that $\pi^{\star}$ does not have a threshold-based structure with respect to the other system state variables, i.e., the levels of batteries and the channel power gains, for the case of $N > 1$. However, for the case of $N = 1$, the following Theorem provides more structural properties of the optimal policy $\pi^{\star}$ with respect to all system state variables.
%\textcolor{blue}{[may be we can provide an intuitive argument to why the optimal policy does not have a threshold-based structure with respect to other state variables]}
\begin{theorem}\label{theorem:2}
Given $N = 1$, for any $s^1 = (b^1_1,A^1_1,g^1_1,h^1_1)$ and $s^2 = (b_1^2,A_1^2,g_1^2,h_1^2)$, the age-optimal policy $\pi^\star$ has the following structural properties: \\(i) When $s^1 \preceq s^2$ and $b_1^1 \geq {\rm max} \left \{b_{\rm max,1} - e^{{\rm H},1}_1,e^{{\rm T},1}_1 \right \}$, if $\pi^\star(s^1) = T$, then $\pi^\star(s^2) = T$. \\(ii) When $s^1 \succeq s^2$ and $b^2_1 \geq {\rm max} \left \{b_{{\rm max},1} - e^{{\rm H},2}_1,e^{{\rm T},2}_1 \right \}$, if $\pi^\star(s^1) = H$, then $\pi^\star(s^2) = H$.\\
Note that the symbols $\preceq$ and $\succeq$ represent the element-wise inequalities.
\end{theorem}
\begin{IEEEproof}
%First, we note that proving that $\pi^\star(s_1) = a$ leads to $\pi^\star(s_2) = a$ is equivalent to showing that
%\begin{align}\label{struc_prop}
%Q(s_2,a) - Q(s_2,a') \leq Q(s_1,a) - Q(s_1,a'), \forall a' \neq a,
%\end{align}
%where this holds since if $a$ is optimal in state $s_1$, then we have $Q(s_1,a) - Q(s_1,a') \leq 0, \forall a'$, which leads to $Q(s_2,a) \leq Q(s_2,a'),\forall a'$, i.e., taking action $a$ is optimal in state $s_2$. Hence, 
Since the action space becomes $\mathcal{A} \triangleq \{H,T_1\} $ for the case of $N= 1$, (i) is proven ((ii) is proven) if (\ref{struc_prop}) holds for $\bar{a} = T_1$ and $a' = H$ ($\bar{a} = H$ and $a' = T_1$). Therefore, in the remaining, we focus on the proof of (i) while (ii) can be proven similarly. Particularly, from (\ref{transprob})-(\ref{eq:AoI_transprob}) and (\ref{Q_func}), we have
\begin{align}\label{Q_Si,T}
Q(s^n,T_1) = A^n_1 + C_0\sum\limits_{g'_1} \sum\limits_{h'_1}{ V(b^n_1 - e^{{\rm T},n}_1,1,g'_1,h'_1)},
\end{align}
\begin{align}\label{Q_Si,H}
Q(s^n,H) = A_1^n + C_0\sum\limits_{g'_1} \sum\limits_{h'_1}{ V(b_{{\rm max},1},{\rm min}\{A_{{\rm max},1}, A_1^n + 1 \},g'_1,h'_1)},
\end{align} 
where $n \in \{1,2\}$ and the next battery level in (\ref{Q_Si,H}) is equal to $b_{{\rm max},1}$ since $b_1^1 + e^{{\rm H},1}_1 \geq b_{{\rm max},1}$ and $b_1^1 \leq b_1^2$. Since $s^1 \preceq s^2$ and based on Lemma \ref{Lemma:2}, we have $V(b^1_1 - e^{{\rm T},1}_1,1,g'_1,h'_1) \geq V(b^2_1 - e^{{\rm T},2}_1,1,g'_1,h'_1)$ ($e^{{\rm T},1}_1 \geq e^{{\rm T},2}_1$) and $V(b_{{\rm max},1},{\rm min}\{A_{{\rm max},1}, A_1^2 + 1 \},g'_1,h'_1) \geq V(b_{{\rm max},1},{\rm min}\{A_{{\rm max},1}, A_1^1 + 1 \},g'_1,h'_1)$. Hence, (\ref{struc_prop}) holds for $\bar{a} = T_1$ and $a' = H$, which completes the proof of (i).
\end{IEEEproof}

\begin{remark}\label{rem:1}
Note that according to Theorem \ref{theorem:2}, the age-optimal policy $\pi^\star$ has a threshold-based structure over the set of states $\mathcal{S}_{\rm d}^{{\rm th},a} \triangleq \left \{s \in \mathcal{S}_{\rm d}^{a}: b_1 \geq {\rm max} \{b_{{\rm max},1} - e^{\rm H}_1, e^{\rm T}_1 \}\right \}$, for the case of $N = 1$. Particularly, $\pi^\star$ is a threshold-based policy with respect to each of the system state variables, i.e., $b_1,A_1,g_1,$ and $h_1$. For instance, for a fixed $(b_1,g_1,h_1)$, if $A_{{\rm th},1}$ is the minimum value of AoI for which it is optimal to take an action $a = T_1$, then for all states $s \in \mathcal{S}_{\rm d}^{\rm th}$ such that $A_1 \geq A_{{\rm th},1}$, the optimal decision is $T_1$ as well. In addition, if there exists a state $s^{\rm th} = (b_{{\rm th},1},A_{{\rm th},1},g_{{\rm th},1},h_{{\rm th},1})$, where $b_{{\rm th},1},g_{{\rm th},1}$, and $h_{{\rm th},1}$ are defined similar to $A_{{\rm th},1}$, then $\pi^\star(s) = T_1, \forall s \in \mathcal{S}_{\rm d}^{\rm th}$, such that $s \succeq s^{\rm th}$.
\end{remark}

It is worth noting that the case of $N = 1$ in our system setup refers to the classical single source-destination pair model studied in most prior works on AoI in the literature, e.g., \cite{kaul2012real,kam2013age,costa2016age,
chen2016age,Basel,kosta2017age,
javani2019age,sun2017update}. Since the single source-destination pair model may actually be sufficient to study a diverse set of applications \cite{kaul2012real} (e.g., predicting and controlling forest fires, safety of an intelligent transportation system, and efficient energy utilization in future smart homes), the results obtained in Theorem \ref{theorem:2} for $N=1$ are of interest on their own in many applications. Furthermore, the results of Theorem \ref{theorem:2} are very useful to investigate the differences between the structural properties of the age-optimal and throughput-optimal policies for the single source-destination pair model, as will be discussed in the next section.
%This is also intuitive, since when the value of AoI becomes large, it is optimal to update the status of information at the destination by sending a new update packet.
\section{Age-optimal Policy vs. Throughput-optimal Policy}\label{sec: comparison}
In this section, we aim to analytically compare the structural properties of the age-optimal and the throughput-optimal policies. Due to its higher tractability (as demonstrated in the previous section), we will focus on the single source-destination pair model for this comparison. Specifically, we first formulate the average throughput maximization problem for the case of $N = 1$ in the system setup presented in Section \ref{sec:Model}. Afterwards, we investigate some structural properties of the throughput-optimal policy from which we highlight the differences between the structures of the age-optimal and throughput-optimal polices.
\subsection{Average Throughput Maximization Formulation and Proposed Solution}
When the objective is to maximize the average throughput, the system state at slot $k$ for the case of $N = 1$ is defined as $s(k) = \left \{b_1(k), g_1(k), h_1(k) \right \} \in \mathcal{S}_{\rm d}^{r}$, where $\mathcal{S}_{\rm d}^{r}$ is the state space of the discrete model for the throughput maximization problem, i.e., when the battery and channel power gain are discretized. Note that the AoI is not included now in the state of the system. For such single source-destination pair model, the action space is defined as $\mathcal{A} \triangleq \{H,T_1\} $, where the source node can either harvest energy or transmit a packet of size $S$ at each time slot. The evolution of the battery is then given by $(\ref{eq:batt_evol_dis})$. Hence, the average throughput maximization problem is modeled as a finite-state finite-action MDP for which there exists an optimal stationary deterministic policy \cite{bertsekas2011dynamic}. Particularly, under a policy $\mu$, the long-term average throughput is defined as
\begin{align}\label{average_throughput}
\bar{R}^{\mu}_1 \triangleq \liminf_{K\to \infty} \frac{1}{K+1} \sum_{k = 0}^K \mathbb{E}\left[\mathbbm{1}\left(a(k) = T_1 \right) S\g s(0)\right], 
\end{align}
where the system receives some reward equal to $S$ in an arbitrary time slot only if this slot is allocated for data transmission to the destination node. Our goal is then to characterize the throughput-optimal policy $\mu^*$ which maximizes the long-term average throughput such that 
\begin{align}\label{Optim_prob_throughput}
\mu^\star = {\rm arg}\; \underset{\mu}{\rm max} \;\bar{R}^{\mu}_1.
\end{align}

Under a stationary deterministic policy $\mu$, the probability of moving from state $s$ to state $s'$ can be expressed as
\begin{align}\label{transprob_throughput}
\P \left(s' \g s, \mu(s)\right) = \P \left(b'_1 \g b_1, g_1, h_1, \mu(s)\right) \P(g'_1) \P(h'_1),
\end{align}
where $\P \left(b'_1 \g b_1, g_1, h_1, \mu(s)\right)$ can be expressed as in (\ref{eq:batt_transprob}). The optimal policy $\mu^\star$ can then be obtained by solving the following Bellman's equation using the VIA (similar to (\ref{value_function_itern}) and (\ref{policy_itern}))
\begin{align}\label{belman_equation_throughput}
\bar{R}^\star + V(s) = \underset{a \in \mathcal{A}(s)}{\rm max} Q(s,a), s \in \mathcal{S}_{\rm d}^r,
\end{align}
where $\bar{R}^\star$ is the optimal average throughput achievable by $\mu^\star$ and $Q(s,a)$ can be expressed as
\begin{align}\label{Q_func_throughput}
Q(s,a) = \mathbbm{1}\left(a = T_1 \right) S + \sum\limits_{s' \in \mathcal{S}_{\rm d}^r} {\P(s' \g s, a) V(s')},
\end{align}
where $\P(s' \g s, a)$ is computed by (\ref{transprob_throughput}) and $\mu^\star(s)$ is given by
\begin{align}\label{eq:optimal_policy_throughput}
\mu^\star(s) = {\rm arg}\; \underset{a \in \mathcal{A}(s)}{\rm max} Q(s,a).
\end{align}
%In the following, we first characterize some structural properties of the throughput-optimal policy $\mu^\star$ using which we compare the structures of the age-optimal and throughput-optimal policies. 
%The value function $V(s)$ can be evaluated iteratively using the VIA[]. Particularly, according to the VIA, the value function at iteration $m$, $m = 1, 2, \cdots$, is computed as 
%\begin{align}\label{value_function_itern}
%\nonumber V(s)^{(m)} & = \underset{a \in \mathcal{A}(s)}{\rm min} Q(s,a)^{(m - 1)} \\
%&= \underset{a \in \mathcal{A}(s)}{\rm min} \left \{A(s) + \sum\limits_{s' \in \mathcal{S}^a_{\rm d}} {\P(s' \g s, a) V(s')^{(m - 1)}}\right \},
%\end{align}
%where $s \in \mathcal{S}_{\rm d}^a$. Hence, the optimal policy at iteration $m$ is given by
%\begin{align}
%\pi^{\star(m)}(s) = {\rm arg} \underset{a \in \mathcal{A}(s)}{\rm min} Q(s,a)^{(m - 1)}.
%\end{align}
\subsection{Structural Properties of the Throughput-optimal Policy}
\begin{lemma}\label{Lemma:4}
The value function $V(b_1,g_1,h_1)$, corresponding to the throughput-optimal policy $\mu^\star$, is non-decreasing with respect to the battery level $b_1$, the downlink channel power gain $g_1$, and the uplink channel power gain $h_1$.
\end{lemma}
\begin{IEEEproof}
By using (\ref{transprob_throughput}), the result can be obtained using the same approach used in the proof of Lemma \ref{Lemma:2}, i.e., by applying mathematical induction to the iterations of the VIA.
\end{IEEEproof}
Using Lemma \ref{Lemma:4}, some structural properties of the throughput-optimal policy are presented in the following Theorem.

\begin{theorem}\label{theorem:3}
For any $s^1 = (b_1^1,g_1^1,h_1^1)$ and $s^2 = (b^2_1,g^2_1,h^2_1)$, the throughput-optimal policy $\mu^\star$ has the following structural properties: \\(i) When $s^1 \preceq s^2$ and $b^1_1 \geq {\rm max} \left \{b_{{\rm max},1} - e^{{\rm H},1}_1,e^{{\rm T},1}_1 \right \}$, if $\mu^\star(s^1) = T_1$, then $\mu^\star(s^2) = T_1$. \\(ii) When $s^1\succeq s^2$ and $b^2_1 \geq {\rm max} \left \{b_{{\rm max},1} - e^{{\rm H},2}_1,e^{{\rm T},2}_1 \right \}$, if $\mu^\star(s^1) = H$, then $\mu^\star(s^2) = H$.
\end{theorem}
\begin{IEEEproof}
This result can be obtained using the same approach used in the proof of Theorem \ref{theorem:2}. Note that since this is a maximization problem, proving that $\mu^\star(s^1) = \bar{a}$ leads to $\mu^\star(s^2) = \bar{a}$ is now equivalent to showing that
\begin{align}\label{struc_prop_throughput}
Q(s^2,\bar{a}) - Q(s^2,a') \geq Q(s^1,\bar{a}) - Q(s^1,a'), \forall a' \neq \bar{a}.
\end{align}
\end{IEEEproof}

\begin{remark}\label{Remark:2}
Similar to Remark \ref{rem:1}, Theorem {\ref{theorem:3}} shows that the throughput-optimal policy has a threshold-based structure over the set of states $\mathcal{S}_{\rm d}^{{\rm th},r} = \left \{s \in \mathcal{S}_{\rm d}^{r}: b_1 \geq {\rm max} \{b_{{\rm max},1} - e^{\rm H}_1, e^{\rm T}_1 \}\right\}$. 
\end{remark}
\begin{remark}\label{Remark:3}
Our results in Theorems \ref{theorem:2} and \ref{theorem:3} clearly demonstrate that the structures of the age-optimal and throughput-optimal policies are different, which will also be verified in the numerical results section. Specifically, let us consider a state $\bar{s} = (\bar{b}_1,\bar{g}_1,\bar{h}_1) \in \mathcal{S}_{\rm d}^{{\rm th},r}$ such that $\mu^\star(\bar{s}) = T_1$. Note that the set of states $\mathcal{\bar{S}}_{\rm d}^{{\rm th},a} = \left \{(b_1,A_1,g_1,h_1): (b_1,g_1,h_1) = \bar{s}, 1 \leq A_1 \leq A_{{\rm max},1}\right \}$ belongs to $\mathcal{S}_{\rm d}^{{\rm th},a}$ since $s \in \mathcal{S}_{\rm d}^{{\rm th},r}$. Similar to the definition of $A_{{\rm th},1}$ in Remark \ref{rem:1}, let us define $\bar{A}_{{\rm th},1} = {\rm min}\left(\left \{A_1: \pi^{\star}(\bar{b}_1,A_1,\bar{g}_1,\bar{h}_1 = T_1)\right\}\right)$. Now, for a given state $s \in \mathcal{\bar{S}}_{\rm d}^{{\rm th},a}$ such that $A_1 < \bar{A}_{{\rm th},1}$, according to Lemma \ref{theorem:3}, we note that $\pi^\star(s) = H$. This indicates that $\mu^\star(\bar{s})$ and $\pi^\star(s)$ are different even though the states $s$ and $\bar{s}$ have the same combination $(\bar{b}_1,\bar{g}_1,\bar{h}_1)$ which demonstrates the difference between the structures of the age-optimal and the throughput-optimal polices.
%Specifically according to the age-optimal policy, it is not optimal to allocate a time slot for update packet transmission when the AoI is small since it is better to save the energy in the battery for future update packet transmissions when the AoI becomes large. On the other hand, it may be optimal to transmit an update packet for throughput maximization even if the AoI at the destination is small.
\end{remark}
\section{Numerical Results}\label{sec:results}
In this section, we verify our analytical results derived in section \ref{sec:analyt}, and show the performance of our proposed DRL algorithm in terms of the achievable average weighted sum-AoI as a function of system design parameters. The downlink and uplink channel power gains between the destination and source nodes are modeled as $g_i = h_i = \Gamma \psi^2 d_i^{-\nu}$; where $\Gamma$ is the signal power gain at a reference distance of $1$ meter, $\psi^2\sim \exp(1)$ denotes the small-scale fading gain, and $d_i^{-\nu}$ represents standard power law path-loss with exponent $\nu$. Recall that we denote the number of discrete values that the state variables $g_i$ and $h_i$ can take by $G_i$ and $H_i$, respectively. In the following, we use $g_i = j$ $(h_i = j)$ to refer to the value of the channel power gain at its $j$-th level where $j \in \{1,2,\cdots,G_i\}$ ($j \in \{1,2,\cdots,H_i\}$). Unless otherwise specified, we use the following values for different system parameters: $W = 1$ MHz, $P = 37$ dBm, $\eta = 0.5$, $\sigma^2 = - 95$ dBm, $\Gamma = 0.2$, $\nu = 2$ and $\theta_i = \frac{1}{N}, \forall i \in \mathcal{I}$.
%$d = 25$ meters, $S = 12$ Mbits, $B_{\rm max} = 0.3$ mjoules, $A_{\rm max} = 10$ and  $b_{\rm max} = 9$.
%  It should be noted that the time slot and bandwidth values were normalized to unity to maintain generality (since this example is not specific to any particular technology).
\subsection{Verification of Analytical Results}
\begin{figure*}[t!]
\centerline{
\subfloat[]{\includegraphics[width=0.5\textwidth]{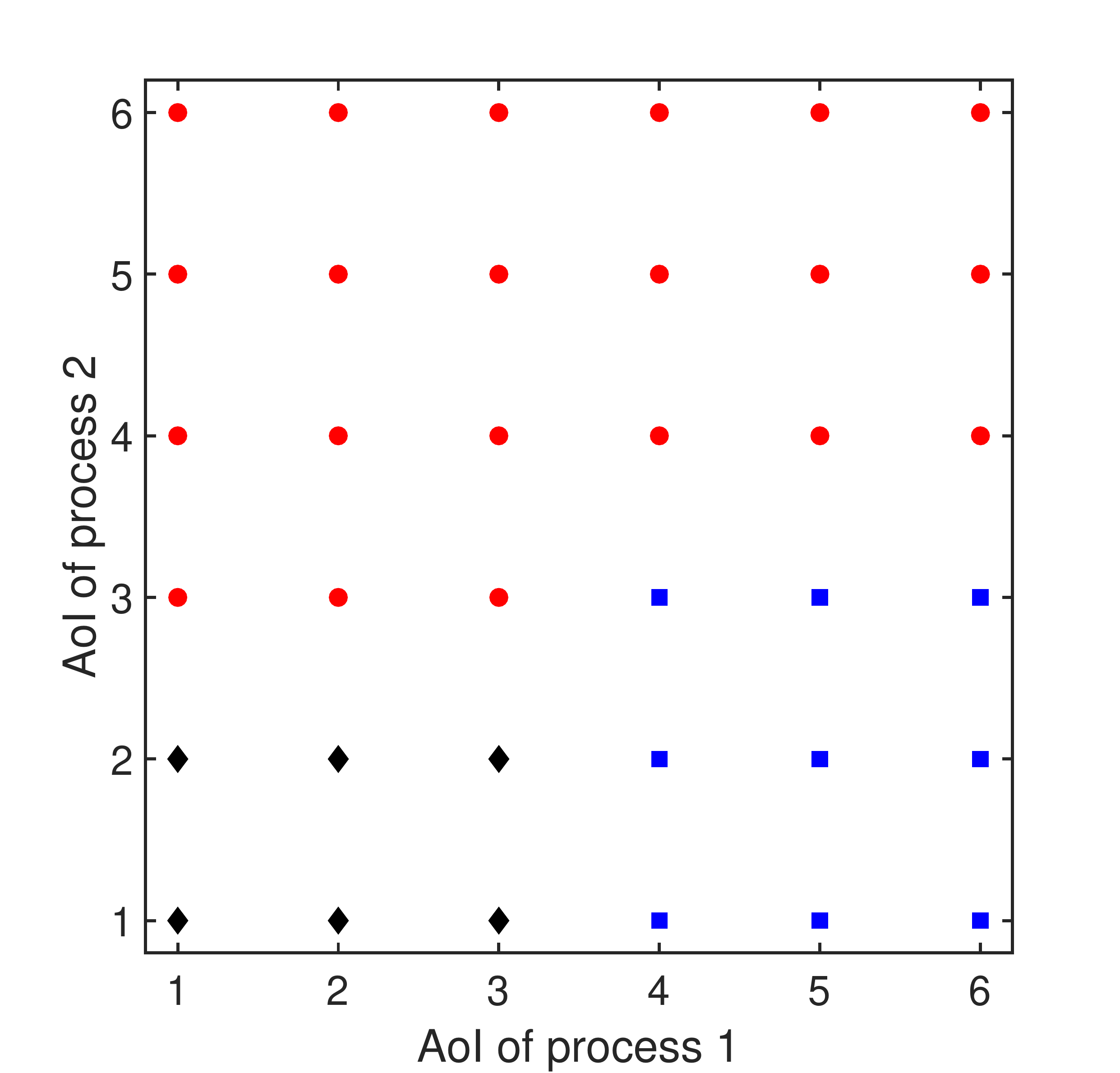}%
\label{f:AoI_structure_2_a}} \hfil
\subfloat[]{\includegraphics[width=0.5\textwidth]{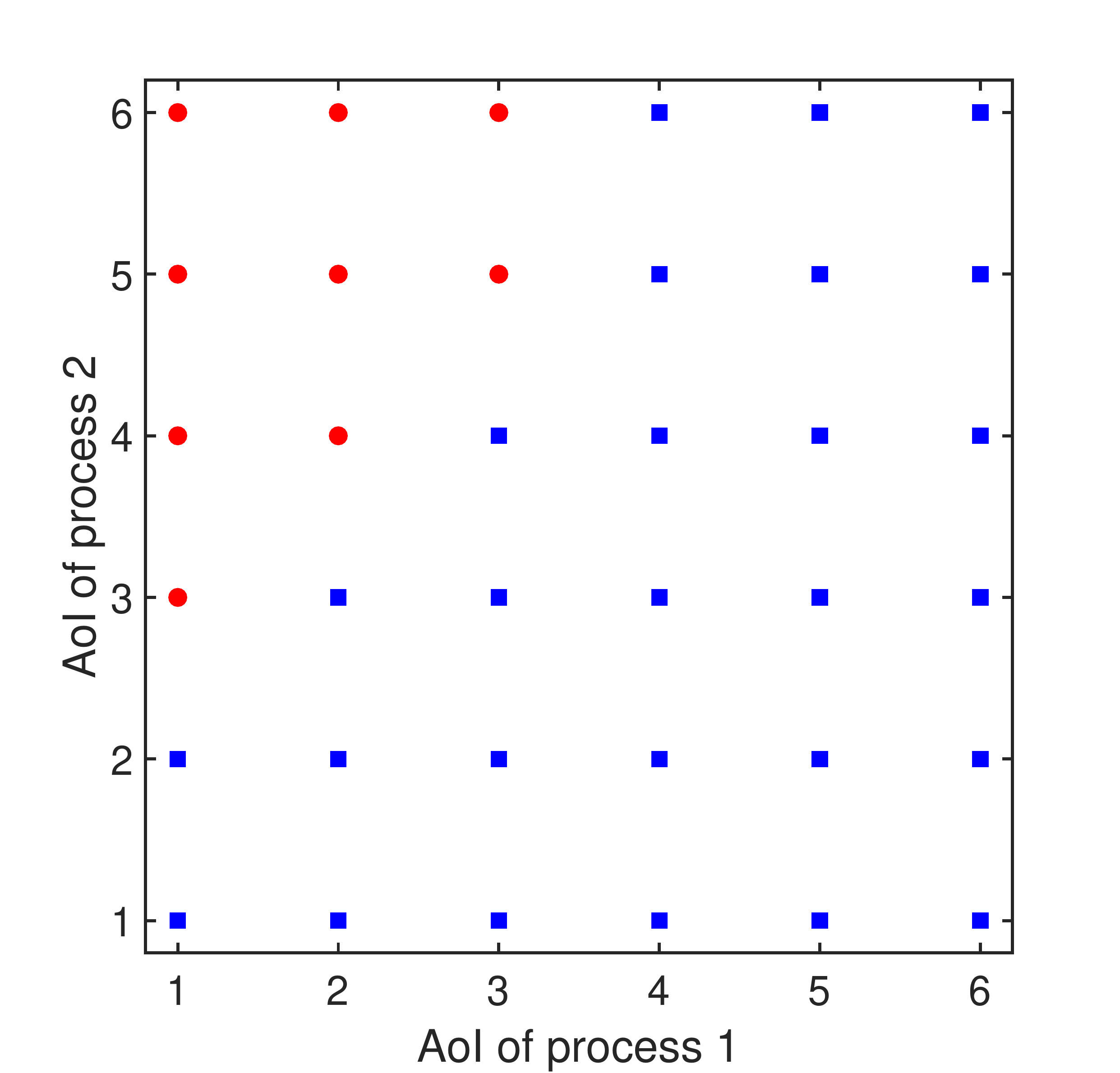}%
\label{f:AoI_structure_2_b}} } \caption{Structure of the age-optimal policy when $N = 2$: (a) $b_1 = b_2 = 1$ and $g_1 = g_2 = 6$, and (b) $b_1 = b_2 = 5$ and $g_1 = g_2 = 2$. We use $d_1 = 25$ meters, $d_2 = 40$ meters, $B_{{\rm max},1} = B_{{\rm max},2} = 0.4$ mjoules, $S = 15$ Mbits, $A_{{\rm max},i} = H_i = G_i = 6, \forall i \in\{1,2\}$ and $b_{{\rm max},1} = b_{{\rm max},2} = 5$.}\label{f:AoI_structure_2}
\end{figure*}

\begin{figure*}[t!]
\centerline{
\subfloat[]{\includegraphics[width=0.5\textwidth]{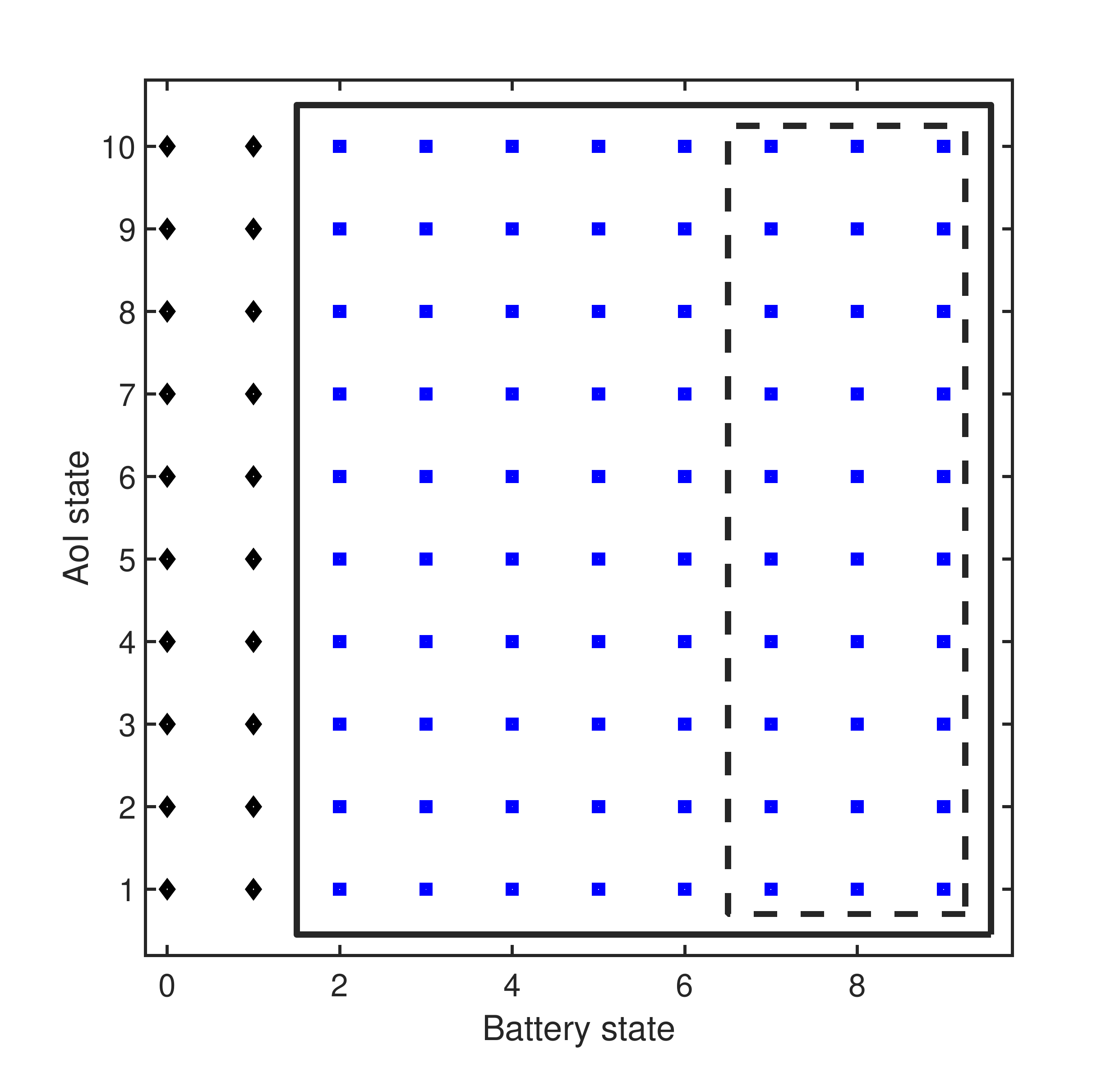}%
\label{f:AoI_structure_1_a}} \hfil
\subfloat[]{\includegraphics[width=0.5\textwidth]{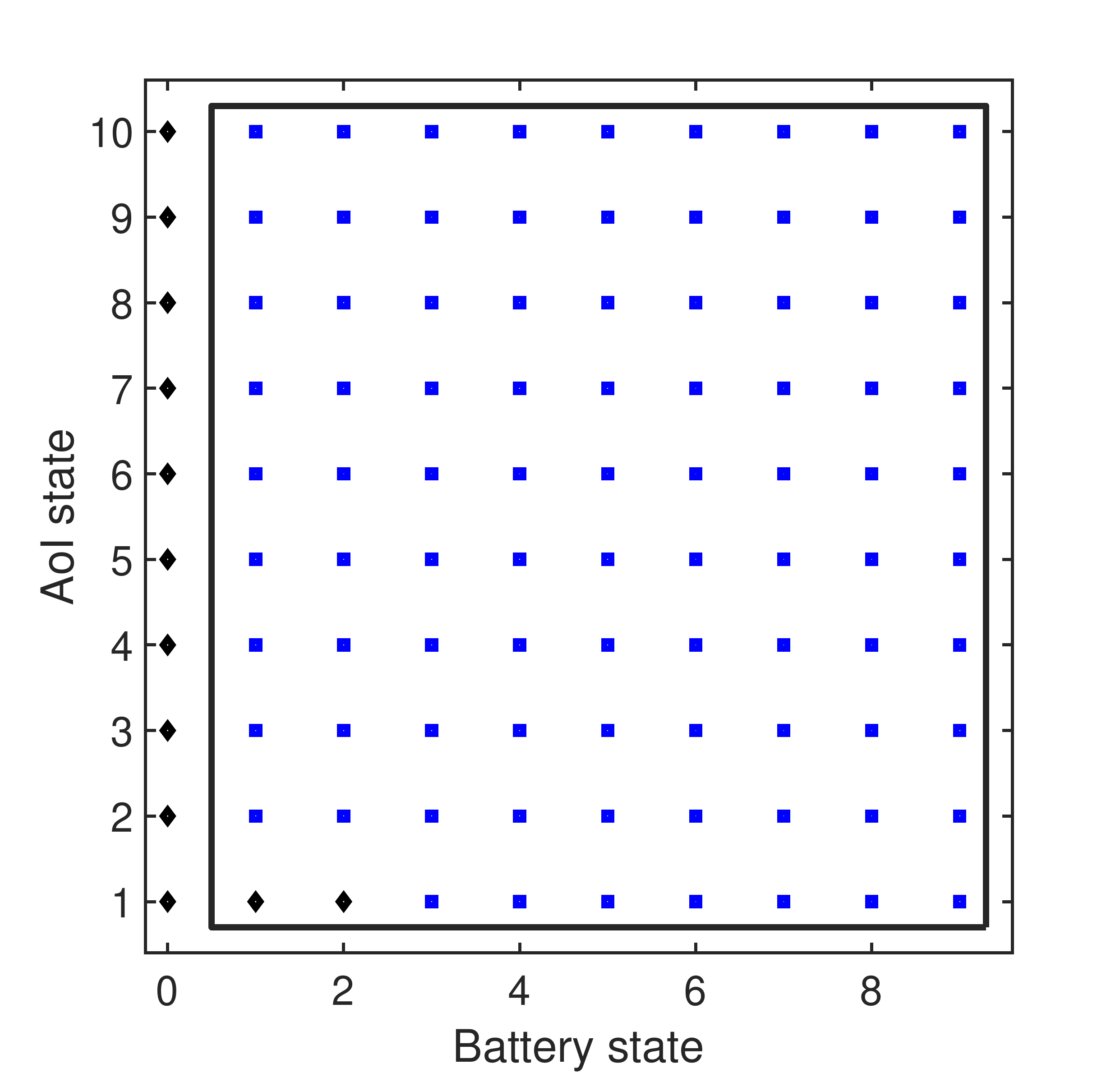}%
\label{f:AoI_structure_1_b}} }
 \caption{Structure of the age-optimal policy when $N = 1$: (a) $g_1 = 2$, and (b) $g_1 \in\{5,6,\cdots,10\}$. We use $d_1 = 35$ meters, $B_{{\rm max},1} = 0.3$ mjoules, $S = 12$ Mbits, $A_{{\rm max},1} = H_1 = G_1 = 10$ and $b_{{\rm max},1} = 9$.}\label{f:AoI_structure_1}
\end{figure*}

In Figs. \ref{f:AoI_structure_2} and \ref{f:AoI_structure_1}, we present the structure of the age-optimal policy for the case of $N= 2$ and $N = 1$, respectively. Particularly, each point in both the figures represents a potential state of the system where a blue square point (a red circle point) (a black diamond point) indicates that the optimal action at this state is $T_1$ ($T_2$) $(H)$. In addition, in Fig. \ref{f:AoI_structure_1}, the points located inside the solid polygon refer to the states for which it is possible to take $T_1$ action, i.e., for each of those states $b_1 \geq e_1^{\rm T}$. Furthermore, the points located inside the dotted polygon represent the set $\mathcal{S}_{\rm d}^{{\rm th},a}$. Note that the dotted polygon is the same as the solid one in Fig. \ref{f:AoI_structure_1_b}. From these results, we can easily verify that the analytical structural properties of the age-optimal policy, derived in Theorems \ref{theorem:1} and \ref{theorem:2}, are satisfied. For instance, in Fig. \ref{f:AoI_structure_2_a}, since the optimal action at the point $(2,3)$ is $T_2$, we observe that the optimal action at the points $(2,y)$, where $y > 3$, is $T_2$ as well (Theorem \ref{theorem:1}). In addition, in Fig. \ref{f:AoI_structure_1_b}, the optimal action at the point $(1,2)$ is $T_1$, and hence, we observe that it is optimal to take action $T_1$ at all the states $(x,y)$ located inside the set $\mathcal{S}_{\rm d}^{{\rm th},a}$ such that $x \geq 1$ and $y \geq 2$ (Theorem \ref{theorem:2}, (i)). On the other hand, we observe that the optimality of taking action $H$ at the point $(2,1)$ implies that it is optimal to take action $H$ at the point $(1,1)$ as well (Theorem \ref{theorem:2}, (ii)).

\begin{figure*}[t!]
\centerline{
\subfloat[]{\includegraphics[width=0.5\textwidth]{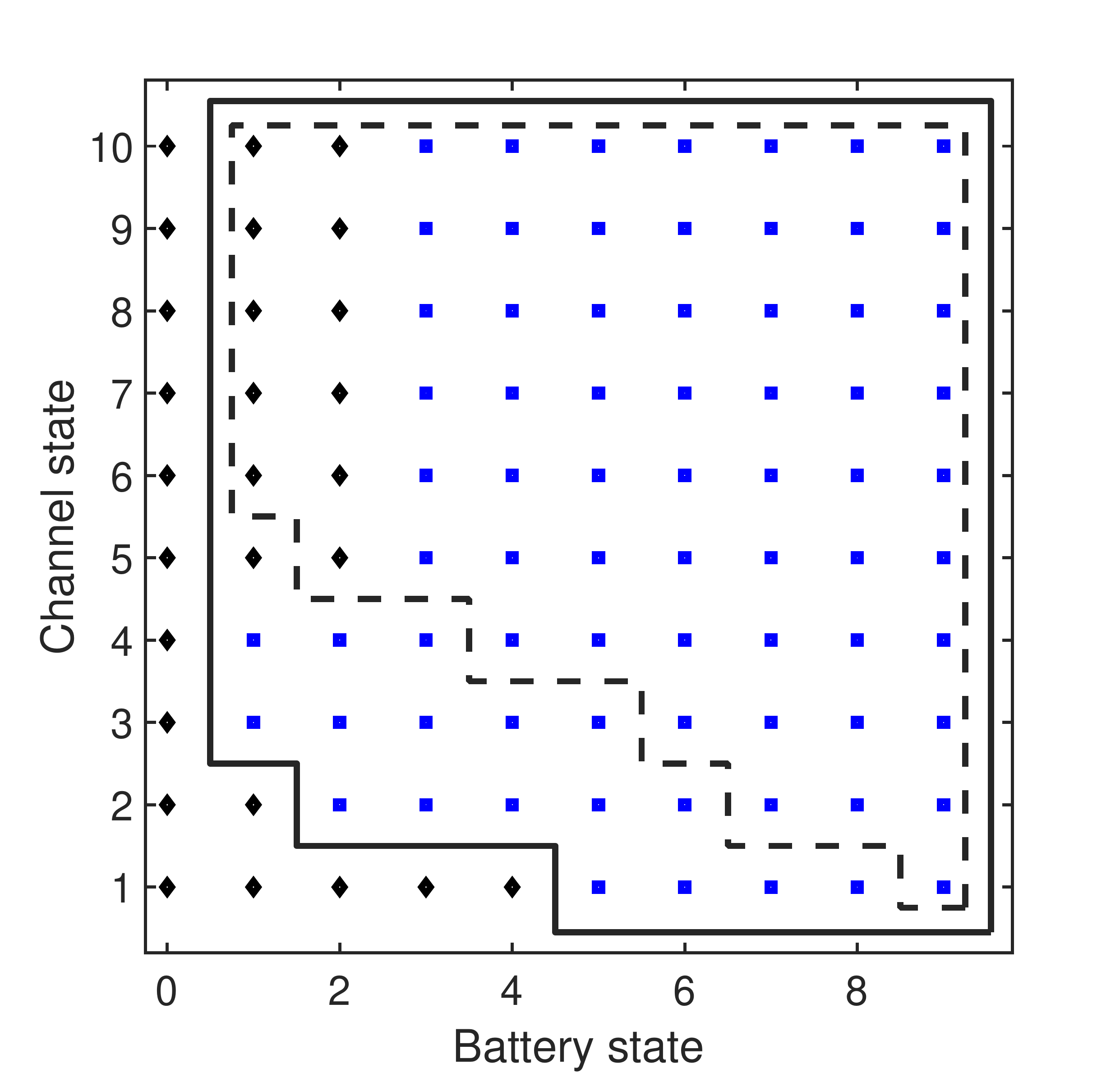}%
\label{f:comb_a}} 
\subfloat[]{\includegraphics[width=0.5\textwidth]{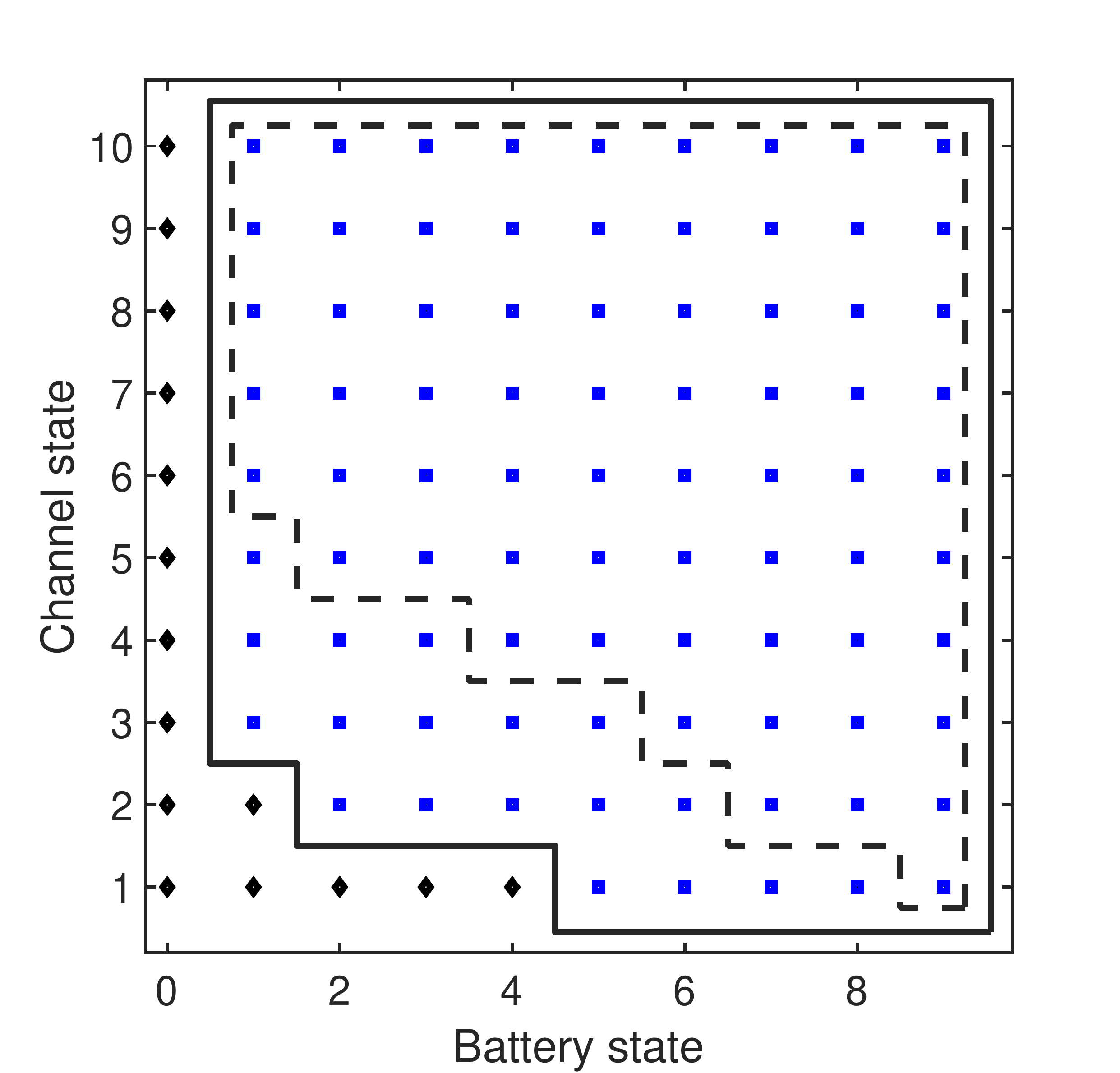}%
\label{f:comb_b}}} \caption{Comparison between the age-optimal and throughput-optimal policies when $N = 1$: (a) Structure of throughput-optimal policy as well as age-optimal policy for $A_1 = 1$, and (b) Structure of age-optimal policy for $A_1 \in \{2,3,\cdots,10\}$. We use the same simulation setup as in Fig \ref{f:AoI_structure_1}.}\label{f:comb}
\end{figure*}    
\subsection{Comparison of the Structures of the Age-optimal and Throughput-optimal Policies}
The difference between the structures of the age-optimal and throughput-optimal polices can be understood by comparing Figs. \ref{f:comb_a} and \ref{f:comb_b}. Specifically, according to the value of $A_1$, we have two different regimes: i) when $A_1$ is small (for instance, $A_1 = 1$ in our simulation setup), the destination node has a fresh information about process $1$, and hence there is no urgency to transmit an update packet, because of which the structures of the age-optimal and throughput-optimal policies are similar (they are the same in our simulation setup when $A_1 = 1$, as shown in Fig. \ref{f:comb_a}), and ii) when $A_1$ is large $(A_1 > 1)$, different from the throughput-optimal policy, it is always optimal to take action $T_1$ regardless of the amount of available energy in the battery according to the age-optimal policy. This is intuitive since if the values of AoI and the battery state are small, it is wise to save the current energy in battery for future update packet transmissions when the AoI value becomes large.

Fig. \ref{f:comb_a} also verifies the analytical structural properties of the throughput-optimal policy, presented in Theorem \ref{theorem:3}. For instance, we observe that it is optimal to take action $T_1$ at all the states $(x,y)$ located inside the set $\mathcal{S}_{\rm d}^{{\rm th},r}$ (i.e., the dotted polygon) such that $x \geq 4$ and $y \geq 4$, since the optimal action at the point $(4,4)$ is $T_1$ (Theorem \ref{theorem:3}, (i)). Furthermore, since the optimal action at the point $(2,10)$ is $H$, we observe that it is optimal to take action $H$ as well at all states $(x,y)$ located inside $\mathcal{S}_{\rm d}^{{\rm th},r}$ such that $x \leq 2$ and $y \leq 10$ (Theorem \ref{theorem:3}, (ii)).

\subsection{Impact of System Design Parameters on Optimal Average Weighted Sum-AoI}
\begin{figure}[t!]
\centering
\includegraphics[width=0.6\columnwidth]{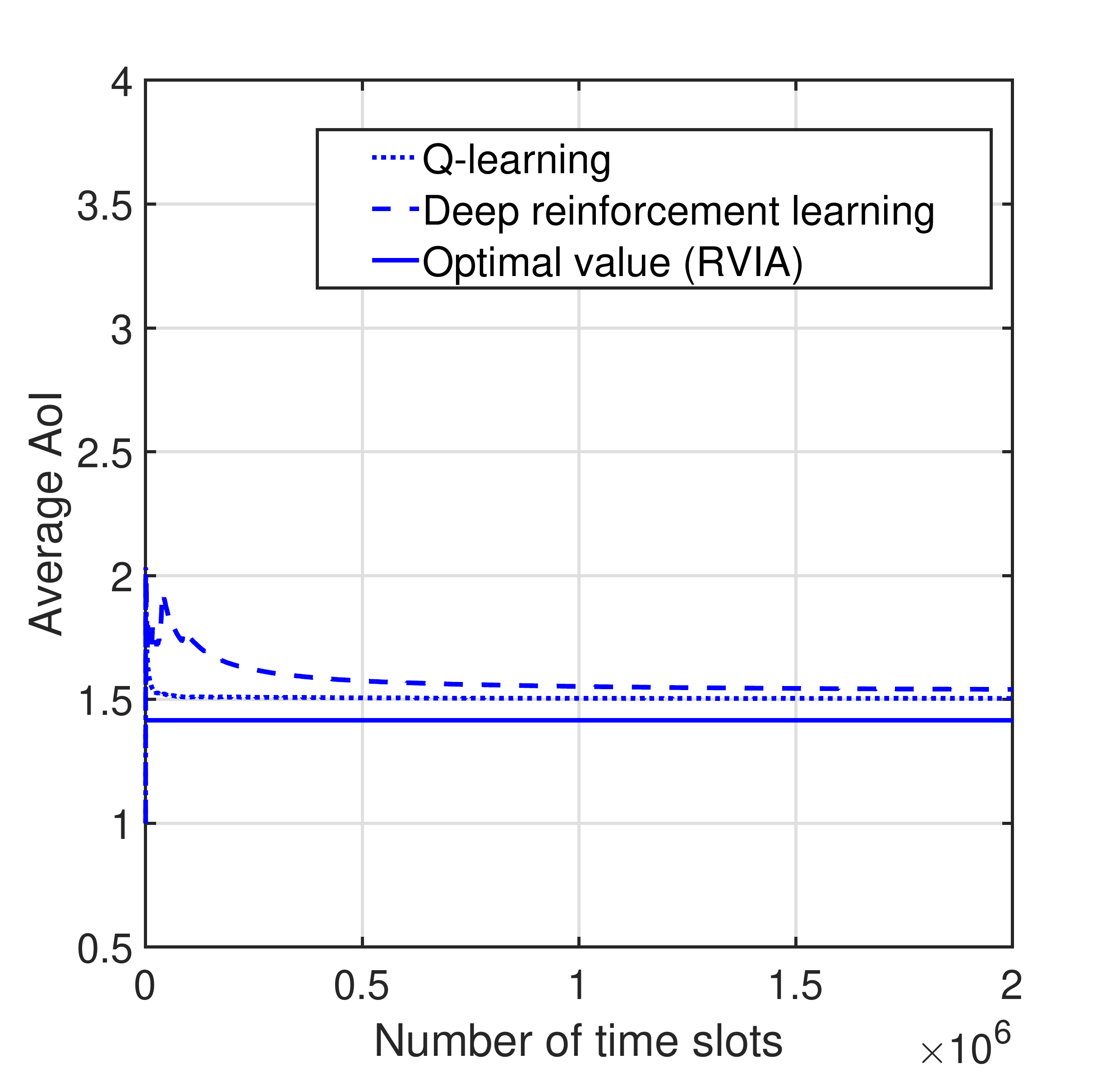}
\caption{Convergence of deep reinforcement learning algorithm when $N = 1$. We use $d_1 = 25$ meters, $B_{{\rm max},1} = 0.3$ mjoules, $S = 12$ Mbits, $A_{{\rm max},1} = H_1 = G_1 = 4$ and $b_{{\rm max},1} = 3$.}
\label{f:conv}
\end{figure}

\begin{figure}[t!]
\centering
\includegraphics[width=0.6\columnwidth]{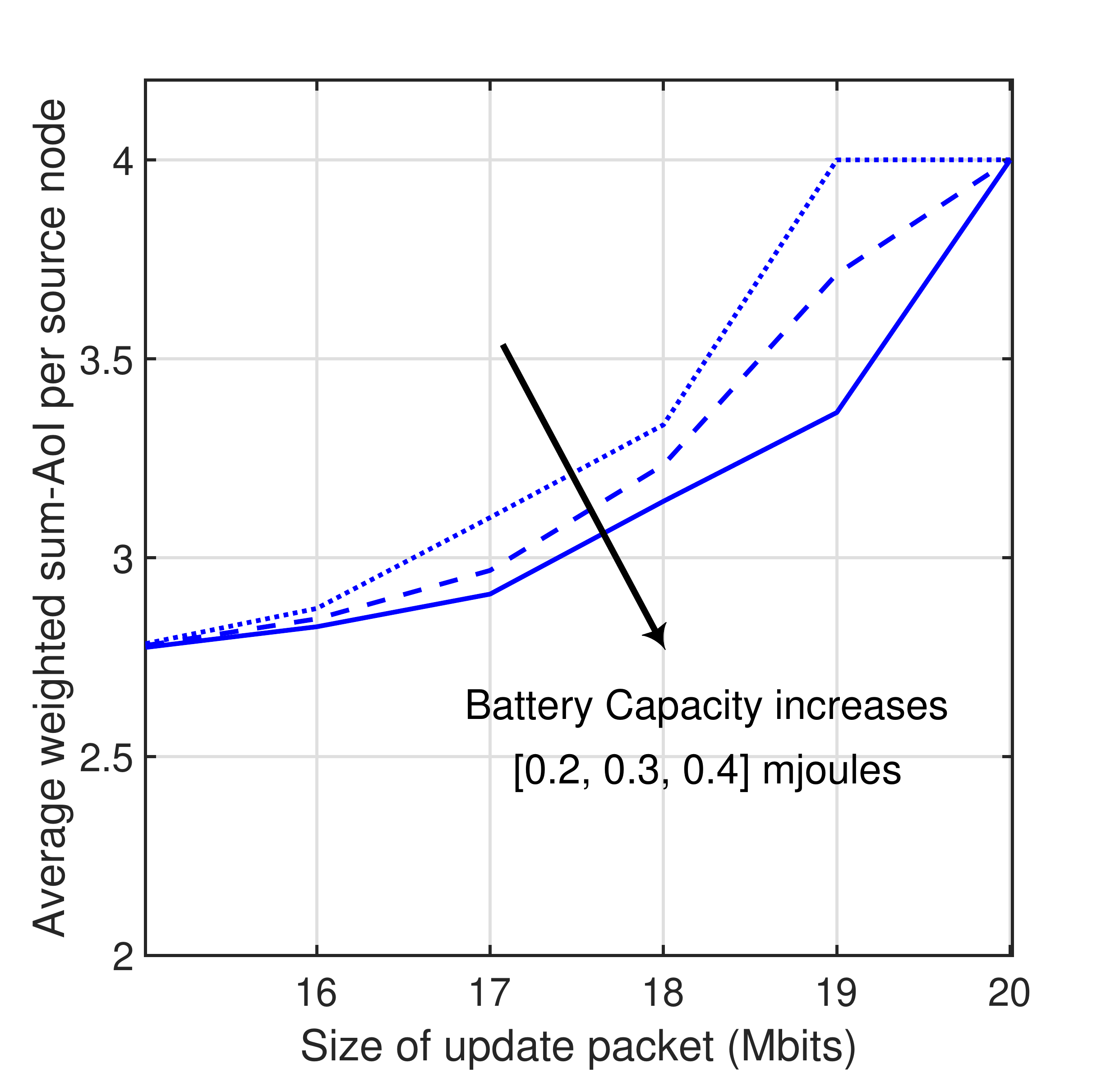}
\caption{Impact of size of update packets and capacity of batteries on the achievable average weighted sum-AoI by the deep reinforcement learning algorithm, for $N = 3$. We use $d_1 = 25$ meters, $d_2 = 40$ meters, $d_3 = 20$ meters, $A_{{\rm max},i} = H_i = G_i = 4, \forall i \in\{1,2,3\}$ and $b_{{\rm max},1} = b_{{\rm max},2} = b_{{\rm max},3} = 3$. We also consider that $B_{{\rm max},1} = B_{{\rm max},2} = B_{{\rm max},3}$.}
\label{f:DRL_perform}
\end{figure}

Due to the curse of dimensionality in the state space of our formulated MDP, the age-optimal policy obtained by applying classical reinforcement learning algorithms \cite{bertsekas2011dynamic}, e.g., the RVIA, can only be evaluated numerically for small-scale settings (i.e., small values for both $N$ and the cardinality of the discrete support set of each state variable). Therefore, we first consider the case of $N = 1$ in Fig. \ref{f:conv} to check the convergence of our proposed DRL algorithm while quantifying its performance in terms of the gap between its achievable average AoI and the optimal value obtained by the RVIA. Afterwards, we demonstrate the impact of system design parameters on the achievable average weighted sum-AoI for a larger value of $N$ $(N = 3)$ in Fig. \ref{f:DRL_perform}, using the DRL algorithm. Clearly, Fig. \ref{f:conv} shows that our proposed reinforcement learning algorithm is able to learn the optimal policy quickly, and hence approaches the optimal average AoI. Note that the slight gap between the optimal value and the achievable average AoI by the DRL algorithm is due to using an $\epsilon$-greedy policy in the DRL algorithm (required for exploring all the state-action pairs while learning the optimal policy, and hence guaranteeing the convergence of the algorithm). However, after the DRL algorithm converges to some value, one can check that the algorithm learns the optimal policy. Hence, the optimal value of average AoI can be achieved by reducing the value of $\epsilon$ to zero after the algorithm has converged (i.e., exploiting the learning process without the need of wasting time in exploring the environment anymore). 
   
 Fig. \ref{f:DRL_perform} shows the impact of the capacity of batteries and size of update packets on the achievable optimal average weighted sum-AoI $\bar{A}^{\star}$, satisfying the Bellman's equations in (\ref{belman_equation}). It is observed that the achievable average sum-AoI monotonically decreases as the size of update packets decreases and/or the capacity of batteries increases. This is due to the fact that decreasing the size of update packets reduces the amount of energy needed to transmit an update packet from each source node, and increasing the capacity of batteries allows to store more harvested energy inside the batteries. This, in turn, increases the likelihood that each source node will have enough energy required for an update packet transmission when the AoI value of its observed process is large, and hence the achievable average weighted sum-AoI is reduced.

%\begin{figure}[t!]
%\centering
%\includegraphics[width=0.8\columnwidth]{throughput_policy.eps}
%\caption{Structure of the throughput-optimal policy.}
%\label{f:through_structure}
%\end{figure}
%
%\begin{figure}
%\centering
%\includegraphics[width=0.8\columnwidth]{S_and_E_impact.eps}
%\caption{Impact of battery capacity and size of update packet on the achievable average AoI.}
%\label{f:EandS_imp}
%\end{figure}
\section{Conclusion}\label{sec:con}
In this paper, we have proposed an implementable age-optimal sampling strategy for designing freshness-aware RF-powered communication systems. In particular, we studied a real-time monitoring system in which multiple RF-powered source nodes are sending update packets to a destination node with the objective of keeping its information status about their observed processes fresh. For this system setup, the long-term average weighted sum-AoI minimization problem was formulated, where the WET by the destination node and scheduling of update packet transmissions from the source nodes are jointly optimized. To obtain the age-optimal policy, the problem was modeled as an average cost MDP with finite state and action spaces. Since the state space in the formulated MDP is extremely large, we proposed a DRL algorithm that can learn the optimal policy efficiently. An analytical characterization for the structural properties of the age-optimal policy was also provided, where it was proven that the age-optimal policy has a threshold-based structure with respect to the AoI values for different processes. Moreover, it was demonstrated that the age-optimal policy has a threshold based structure with respect to all system state variables for the single-source destination pair model. We then extended our analysis to the average throughput maximization problem using which we mathematically characterized key differences in the structural properties of the age-optimal and throughput-optimal policies for our system setup.
%Afterwards, to inspect analytically the differences between the structural properties of the age-optimal and throughput-optimal policies for our system setup, we extended our analysis for the average throughput maximization problem. 

Multiple system design insights were drawn from our numerical results. For instance, they showed that the structures of the age-optimal and throughput-optimal policies in the single source-destination pair model are similar when the AoI value is relatively small (i.e., there is no urgency to update the information status at the destination node). In contrast, the age-optimal and throughput-optimal polices have completely different structures when the AoI value grows. Our results also revealed that the optimal average weighted sum-AoI is a monotonically increasing (decreasing) function with respect to the size of update packets (capacity of batteries at the source nodes).

\bibliographystyle{IEEEtran}
\bibliography{journal_AoI_v0.4}
\end{document}